\documentclass[fleqn,usenatbib]{mnras}
\usepackage{newtxtext,newtxmath}
\usepackage[T1]{fontenc}

\DeclareRobustCommand{\VAN}[3]{#2}
\let\VANthebibliography\thebibliography
\def\thebibliography{\DeclareRobustCommand{\VAN}[3]{##3}\VANthebibliography}

\usepackage{graphicx}	% Including figure files
\usepackage{amsmath}	% Advanced maths commands
\usepackage{float}
\usepackage{subfigure}
\usepackage{soul}       % highlight Text

\title[Thermonuclear X-ray bursts in 4U 1636$-$53]{Insight-HXMT observations of thermonuclear X-ray bursts in 4U 1636$-$53}
\author[\rm Yan et al.]{
Zhe Yan,$^{1,2,3}$%\thanks{E-mail: yanzhe19@mails.ucas.ac.cn}
Guobao Zhang,$^{1,2}$\thanks{E-mail: zhangguobao@ynao.ac.cn}
Yu-Peng Chen,$^{3}$
Shu Zhang,$^{3}$
Mariano Méndez,$^{4}$
Jingqiang Peng,$^{2,3}$
\newauthor
Shuang-Nan Zhang,$^{3}$
Jinlu Qu,$^{3}$
Ming Lyu,$^{5,6}$
Jirong Mao,$^{1}$
Mingyu Ge,$^{3}$
Jiancheng Wang,$^{1,2}$
\\
$^{1}$Yunnan Observatories, Chinese Academy of Sciences, Kunming 650216, People's Republic of China \\
$^{2}$University of Chinese Academy of Science, Beijing 100049, People's Republic of China \\
$^{3}$Key Laboratory of Particle Astrophysics, Institute of High Energy Physics, Chinese Academy of Sciences, Beijing 100049, China \\
$^{4}$Kapteyn Astronomical Institute, University of Groningen, P.O. Box 800, NL-9700 AV Groningen, the Netherlands \\
$^{5}$Department of Physics, Xiangtan University, Xiangtan, Hunan 411105, People’s Republic of China \\
$^{6}$Key Laboratory of Stars and Interstellar Medium, Xiangtan University, Xiangtan, Hunan 411105, People’s Republic of China \\
}

\date{Accepted XXX. Received YYY; in original form ZZZ}

\pubyear{2015}

\begin{document}
\label{firstpage}
\pagerange{\pageref{firstpage}--\pageref{lastpage}}
\maketitle

\begin{abstract}
We conducted an analysis of 45 bursts observed from 4U 1636$-$53.
To investigate the mechanism behind the light curve profiles and the impact of thermonuclear X-ray bursts on the accretion environment in accreting neutron star low-mass X-ray binaries.
This analysis employed both light curve and time-resolved spectroscopy methodologies, with data collected by the \textit{Insight}-HXMT instrument.
We found that 30 bursts exhibited similar light curve profiles and were predominantly in the hard state, and two photospheric radius expansion (PRE) bursts were in the soft state.
The light curves of most bursts did not follow a single exponential decay but displayed a dual-exponential behavior.
The initial exponent had a duration of approximately 6 s.
We utilized both the standard method and the `$f_{\rm a}$' method to fit the burst spectra. 
The majority of the `$f_{\rm a}$' values exceeded 1, indicating an enhancement of the persistent emission during the burst. 
Under the two comptonization components assumption, we suggest that the scattering of burst photons by the inner corona may mainly contribute to the persistent emission enhancement.
We also observed an inverse correlation between the maximum $f_{\rm a}$ and the persistent emission flux in the non-PRE burst.
This anti-correlation suggests that when the accretion rate is lower, there is a greater enhancement of persistent emission during the burst peak.
The prediction based on Poynting-Robertson drag (P-R drag) aligns with this observed anti-correlation. 

\end{abstract}


%%%%%%%%%%%%%%%%%%%%%%%%%%%%%%%%%%%%%%%%%%%%%%%%%%

%%%%%%%%%%%%%%%%% BODY OF PAPER %%%%%%%%%%%%%%%%%%

%-----------------------------------------------------
\section{INTRODUCTION}

Thermonuclear (Type-I) X-ray bursts are sudden increases of the X-ray intensity observed in accreting neutron star low-mass X-ray binaries (NS-LMXBs). 
These bursts are characterized by a rapid rise, within a few seconds to ten seconds, followed by an exponential decay back to the pre-burst level over tens to hundreds of seconds. 
The energy released during a single burst in typically ${10}^{39}-{10}^{41}$ erg \citep{1993SSRv...62..223L,strohmayer_bildsten_2006,2020ApJS..249...32G}.
Such X-ray bursts are generally believed to be caused by unstable thermonuclear burning when the accreted hydrogen and/or helium-rich material accumulates on the surface of the neutron star \citep{2008ApJS..179..360G,2020ApJS..249...32G}.

Type-I X-ray bursts can be categorized into three groups based on their duration.
The first category is the typical one described above.
The second category is "intermediate-duration" bursts, with a duration of about 0.5 hours, which result from thermonuclear burning of pure He layer with a mass of 1$-$2 orders of magnitude higher than that of the typical burst occurring at low accretion rate \citep{2005A&A...441..675I,2006ApJ...646..429C,2008A&A...484...43F}.
The third category is the superburst, lasting approximately 10 hours, which comes from the deeper carbon thermonuclear burning \citep{2000A&A...357L..21C,2001ApJ...559L.127C,2002ApJ...566.1045S}.
Some type-I X-ray bursts reach or slightly exceed the Eddington limit, which can cause the outer atmosphere of the neutron star to expand, resulting in a phenomenon known as PRE bursts \citep{1993SSRv...62..223L}.

The time-resolved energy spectra of type-I X-ray bursts can be used to study the radiation mechanism during the burst.
The radiation during the type-I X-ray bursts is mainly composed of two components, the "burst component" originating from thermonuclear burning, and the "persistent emission" resulting from accretion.
While a single blackbody is commonly employed to describe the radiation of the "burst component", it is still unclear what other components contribute to the radiation during the burst and their origins.
\citet{2013A&A...553A..83I} found it difficult to fit the low and high energy bands of the PRE burst observed by \textit{Chandra} and Rossi X-ray Timing Explorer (\textit{RXTE}) in SAX J$1808.4-3658$ with a single blackbody.
After removing the blackbody component, they observed a similar energy spectrum shape to that before the burst, but the intensity increased.
This implies that the persistent emission from accretion is increased during the burst, while the spectral shape remained unchanged.
Subsequently, \citet{2013ApJ...772...94W,2015ApJ...801...60W} conducted further studies on 1759 bursts from 56 sources, including PRE and non-PRE bursts, using the ‘$f_{\rm a}$’ method considering the change of persistent emission intensity during the burst.
They found that the enhancement of the persistent emission during the burst is a common phenomenon.
It was interpreted as the enhancement of the accretion rate, which is caused by the Poynting-Robertson drag \citep[P-R drag, an inflow of the P-R effect,][]{1989A&A...225...48S,1992ApJ...385..642W,2004ApJ...602L.105B}, this enhancement has an upper limit of $84\%$ and $37\%$ of the Eddington flux in PRE bursts and non-PRE bursts, respectively \citep{2015ApJ...801...60W}.

In addition to the inflow caused by the P-R drag mentioned above, the interaction between the burst and the accretion environment encompasses various factors.
These include the structural changes occurring in the disk and/or corona during the burst, as well as the presence of radiation-driven or thermal outflows \citep{2021ASSL..461..209G}.
The impact of the burst on the disk structure can be known from an "intermediate-duration" burst observed by the Neutron Star Interior Composition Explorer (\textit{NICER}) in IGR J17062$-$6143 in June 2020. 
\citet{2021ApJ...920...59B} found that the burst led to a decrease in X-ray intensity during the cooling stage, dropping below the pre-burst level, which subsequently recovered after three days.
This behavior was attributed to the disturbance caused by the burst to the inner disk. 
Furthermore, numerical simulation conducted by \citet{2020NatAs...4..541F} exploring the effect of the PRE burst on a thin disk revealed two additional interaction effects, alongside radiation-driven outflows.
These effects contribute to the enhanced depletion of the inner disk material during the burst.
The hard X-ray deficit explained by the burst through the Compton scattering cooling has been observed in several sources, such as Aql X$-$1 \citep{2003A&A...399.1151M,2013ApJ...777L...9C}, 4U 1636$-$53 \citep{2018ApJ...864L..30C}, GS 1826$-$238 \citep{2014ApJ...791L..39J}, KS 1731$-$260 \citep{2014A&A...564A..20J}, 4U 1705$-$44 \citep{2014A&A...564A..20J}, and 4U 1728$-$34 \citep{2017A&A...599A..89K}.

% introduce 1636 
Since the detection of the first X-ray burst in the NS-LMXB 4U 1636-53 (V801 Ara) by the 8th Orbiting Solar Observatory (OSO-8) \citep{1976IAUC.3010....1S}, this source has been in persistent activity.
The orbital period of 4U 1636$-$53 is 3.8 hr \citep{1998ApJ...503L.147S}, and the spin frequency inferred by BOs is $\sim$ 581Hz \citep{1997IAUC.6541....1Z,1998ApJ...498L.135S}.
Its companion star mass is $\sim$0.4$M_{\odot}$ under the assumption of a NS mass of 1.4$M_{\odot}$ \citep{2002ApJ...568..279G}.
The distance to this source is 6.0$\pm0.5$ kpc, as determined by its pure helium PRE burst \citep{2018ApJ...857L..24G}.
Recent results of Gaia DR2 indicate that this source is located at a distance 4.4$_{-3.1}^{+1.6}$ kpc \citep{2021MNRAS.502.5455A}.
Consistent with previous research that often utilized a 6 kpc distance, we have also chosen to use a 6 kpc distance in our subsequent analysis to enable comparisons with existing literature.

4U 1636$-$53 is a typical Atoll source, it completes the entire evolution from top right to bottom right along the "C" shaped track in the colour-colour diagram (CCD), and from top left to bottom right in the hardness-intensity diagram (HID).
This evolution corresponds to an increase in the mass accretion rate, transitioning from the low/hard (Island) state to the high/soft (Banana) state \citep{1989A&A...225...48S,1989ESASP.296..203V,2011MNRAS.413.1913Z}.
This entire process spans approximately 30$-$40 days \citep{2005MNRAS.361..602S,2007MNRAS.379..247B}.

Notably, the light curve profile of the burst exhibits diversity in this source, particularly during the decay phase, which cannot be characterized by a single exponential component. 
According to \citet{2017A&A...606A.130I}, the model incorporating the radiative cooling and the rp (rapid proton) process effectively fits the observed burst decay profiles from 1996 to 2012 as observed by \textit{RXTE}.
This implies that heating by the rapid proton capture process (rp process) during the decay phase may lead to the emergence of multi-stage decay phase profiles.
Recently, \citet{2021MNRAS.508.2123R} utilized the `$f_{\rm a}$' method to analyze the spectra observed by \textit{AstroSat}-LAXPC from July 2016 to August 2018. 
They found an enhancement of the persistent emission near the peak of the burst, particularly during the PRE burst. 
The values of $f_{\rm a}$ were found to be in line with those reported by \citet{2013ApJ...772...94W}.
From June 2017 to March 2020 in \textit{NICER}'s observations of this source, \citet{2022ApJ...935..154G} identified pronounced soft excesses in all spectra during the burst by employing the `$f_{\rm a}$' method. 
They interpreted this phenomenon to an enhanced accretion rate induced by the burst through the P-R drag.
\citet{2022A&A...660A..31Z} also found the the evidence of the P-R drag in this source.
The broader energy band advantage of the Hard X-ray Modulation Telescope (\textit{Insight}-HXMT) enables a more comprehensive investigation of the formation mechanism of the light curve profiles and the impact of the bursts on the accretion environment.

In this paper, we report the analysis of type-I X-ray bursts in 4U 1636$-$53 observed by \textit{Insight}-HXMT.
We describe the observations and data analysis in  Section \ref{sec:OBSERVATION AND DATA REDUCTION}, and we present our results in Section \ref{section:RESULTS}.
The implications of our findings are discussed and summarized in Section \ref{sec:DISCUSSION}

%-----------------------------------------------------
%%%-----OBSERVATION------%%%

\section{OBSERVATION AND DATA REDUCTION}
\label{sec:OBSERVATION AND DATA REDUCTION}

\textit{Insight}-HXMT \citep{Zhang2020} is China's first X-ray astronomy telescope.
It contains three main payloads: the High Energy X-ray Telescope (HE, 20–250 keV), the Medium Energy X-ray Telescope (ME, 5–40 keV), and the Low Energy X-ray telescope (LE, 1–12 keV).
We analyze the data using the \textit{Insight}-HXMT Data Analysis software (HXMTDAS) v2.05.
The reduction procedure of the data follows the \textit{Insight}-HXMT data reduction guide v2.05\footnote{http://hxmtweb.ihep.ac.cn/SoftDoc.jhtml}. 
After June 2019, the LE detector temperature went beyond the valid temperature range of the background model, which induced some uncertainty caused by electronic noise under 2 keV, so the extracted light curves and fitted spectrum of the LE data are above 2 keV.
Clean event files are screened by using the good time intervals (GTIs) with the standard criterion: (1) Earth elevation angle >$10^{\circ}$; (2) geomagnetic cutoff rigidity >8 GV; (3) pointing offset angle <$0.04^{\circ}$; (4) and at least 300 s away from the South Atlantic Anomaly.

From 2018 February 6 (MJD 58161) to 2022 July 6 (MJD 59767), a total of 1050 ks of monitoring data were obtained for 4U 1636$-$53 using \textit{Insight}-HXMT.
We identified a total of 70 bursts, with 45 of them having both LE and ME data, and the remaining 25 bursts having only ME data.
The comprehensive information regarding the remaining 25 bursts was unattainable because the primary emission of these bursts occurred in the low-energy band.
As a result, we excluded these bursts from our analysis in this work.
The information of 45 bursts analyzed in this work is shown in Table~\ref{tab:table2}.

\begin{figure} 
	\includegraphics[width=\columnwidth]{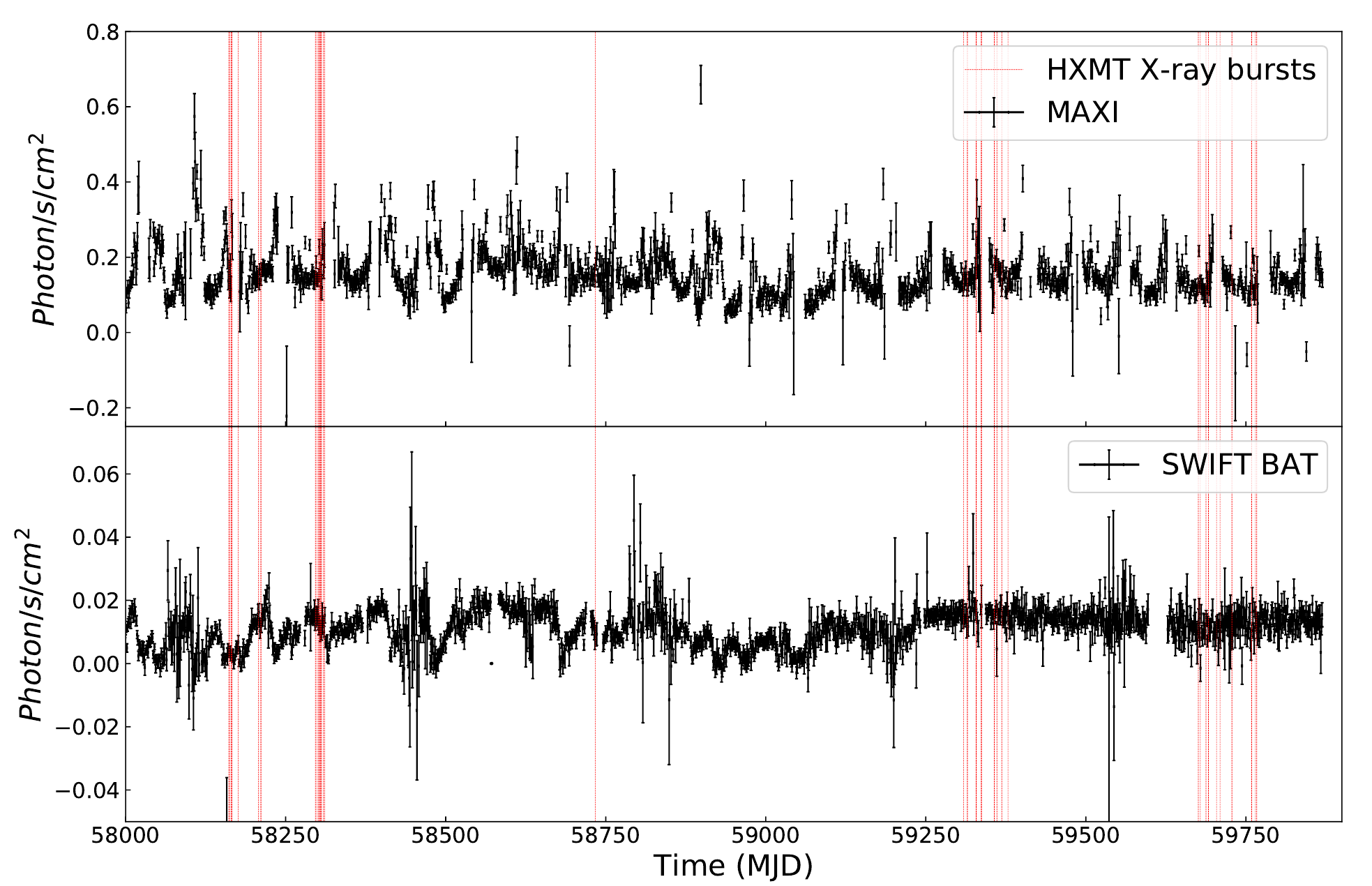}
    \caption{ Top panel: the daily light curve of 4U 1636$-$53 with \textit{MAXI} (2$-$20 keV). Bottom panel: daily light curve of 4U 1636$-$53 with \textit{Swift}-BAT (15$-$50 keV). The 45 bursts observed with \textit{Insight}-HXMT are indicated by red vertical lines.}
    \label{fig:figure1}
\end{figure}

\begin{figure}
	\includegraphics[width=\columnwidth]{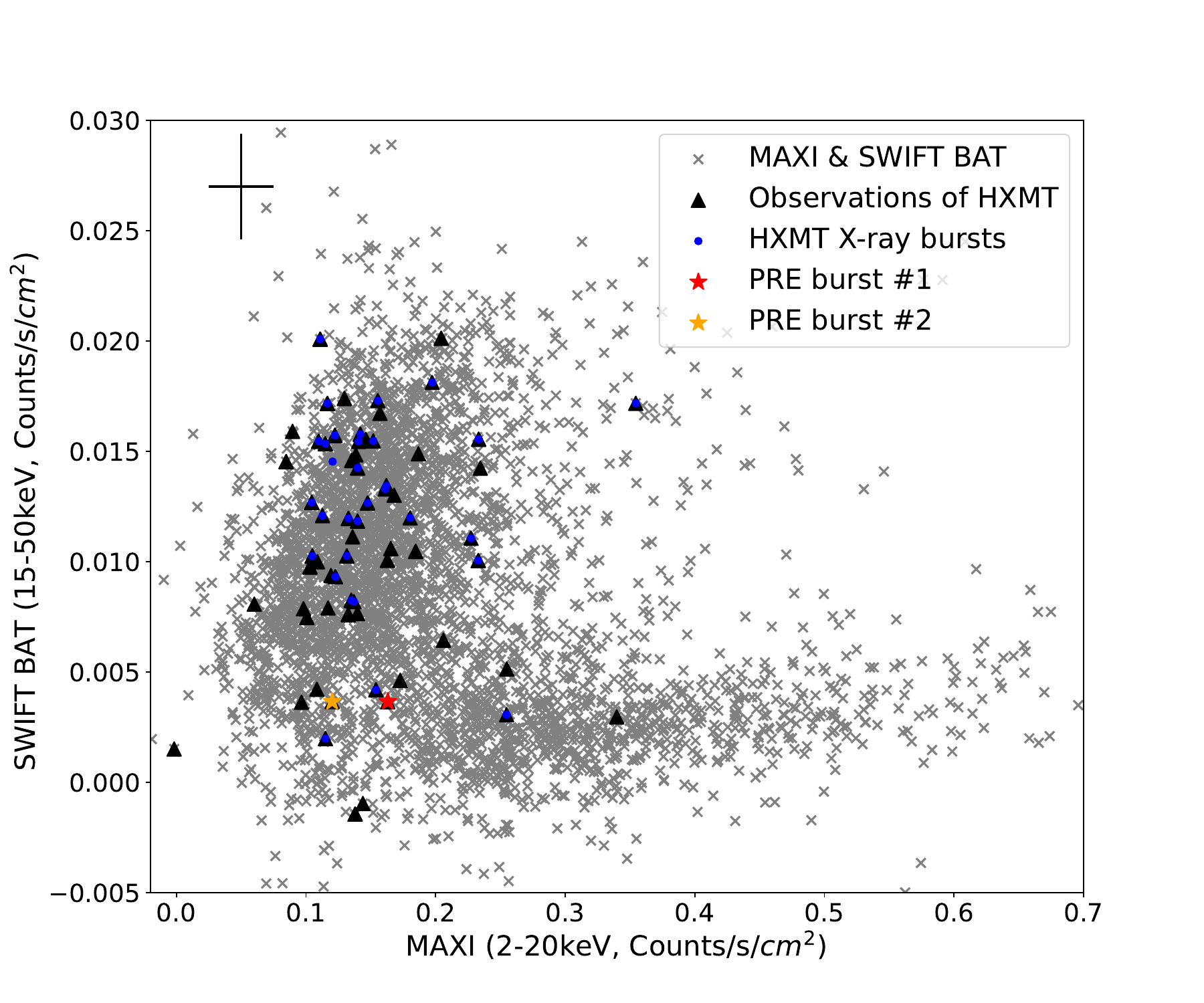}
    \caption{Intensity-intensity diagram of 4U 1636$-$53 with \textit{Swift}/BAT and \textit{MAXI} daily average data.
    The X-axis in the 2$-$20 keV \textit{MAXI} 1-day intensity and the Y-axis in the 15$-$50 keV BAT 1-day intensity.
    The grey 'x' points represent the quasi-simultaneous (the MJD time difference is less than 1 day) observation points of \textit{Swift}-BAT and MAXI.
    The black triangles represent the quasi-simultaneous observation points of \textit{Insight}-HXMT, \textit{Swift}-BAT, and MAXI.
    The black cross at the top left shows the average error bar on each point.
    The locations of 39 out of the detected 45 X-ray bursts are indicated by blue solid points, and the other 6 bursts have no quasi-simultaneous observation data.
    The red and orange stars represent PRE bursts \#1 and \#2, respectively.}
    \label{fig:figure2}
\end{figure}

To illustrate the evolution of the source, we used the daily simultaneous observation data of the Monitor of All-sky X-ray Image (\textit{MAXI}, 2$-$20 keV) and \textit{Swift}/Burst Alert Telescope (BAT, 15$-$50 keV) to make Figure ~\ref{fig:figure1}, and marked the corresponding bursts with red vertical lines.
As the survey with \textit{Insight}-HXMT does not cover the complete state evolution of 4U 1636$-$53, we utilized the observation of \textit{MAXI} and \textit{Swift}-BAT to indicate the burst locations in the evolutionary phase, as depicted in Figure ~\ref{fig:figure2}.

%lc part
To search for type-I X-ray bursts in each observation, we extracted light curves with a time bin of 1 s in the 2$-$10, 10$-$35, and 27$-$250 keV bands for LE, ME, and HE, respectively.
The light curves we used had the instrument background subtracted and went correct by dead time.
We also extracted light curves at 0.25 s to analyze the characteristics of the burst profile.
To determine the start time of each burst in detail, we applied the method described by \citet{2021ApJ...907...79B}.
Firstly, we defined the time of a random point in 0.25s bin size net light curve in a single observation as $t_n$.
We calculated the mean count rate of [$t_{n-30},t_{n-10}$], and roughly defined the bin of $t_n$ as onset time whose count rate exceeded two times the mean count rate.
Next, we precisely searched the onset by identifying the intersection between the first-order Savitzky–Golay filter \citep{1964AnaCh..36.1627S} smoothed light curve (after subtracting the mean count rate) and the 10 percent of the mean peak count rate, which was considered the start time point.
Once we obtained the start time of the burst, the rise time of the burst was defined as the time interval between the start time and the point where the count rate reached one standard deviation of the respective burst peak count rate in the rising phase.
These are shown in Table~\ref{tab:table2}.

%spec part
We extracted the 64-s interval spectrum before the burst as the pre-burst emission (including the persistent emission and the instrumental background), which was taken as the background during bursts.
In cases where there was insufficient time to extract the 64-s spectrum, we utilized the last burst-free part of the same observation.
For the bursts outside the GTI, we used the 64-s spectrum within the GTI to represent the persistent emission in the same observation.
The instrument background and response matrix files for each detector were generated using the established criteria in the \textit{Insight}-HXMT pipeline.
% gen burst spec
To enhance the statistical accuracy of time-resolved spectra, we divided the spectra during the burst into three segments: from the beginning of the burst to half of the peak count rate in the decay phase, from half of the peak count rate to a quarter of the peak count rate, and from a quarter of the peak count rate to the end of the burst.
For bursts with a duration of less than 30 s, we extracted the time-resolved spectra of the previous three parts in a bin size of 1s, 2s, and 3s, respectively.
For bursts lasting more than 30 s, we used bin sizes of 3s, 4s, and 5s for the corresponding parts of the burst.

The spectra during the burst were rebinned using the ftool $\sc grppha$ with a minimum of 20 counts per grouped bin.
However, bursts \#6 and \#14 were too weak, so we used a binning of 10 counts per group.
We added a systematic uncertainty of 1 percent, 2 percent, and 1 percent to the LE, ME, and HE spectra, respectively, based on the information provided by the \textit{Insight}-HXMT team \citep{2020JHEAp..27...64L}.
XSPEC, V12.12.0\footnote{https://heasarc.gsfc.nasa.gov/docs/xanadu/xspec/index.html} was utilized  for spectral analysis.
During the analysis of LE, ME, and HE, the energy bands are chosen to be 2$-$10 keV, 8$-$35 keV, and 25$-$100 keV, respectively.
The errors of parameter values were calculated at the $1\sigma$ confidence level.
In the spectrum fitting, we used the $\sc TBabs$ model with the photoionization cross-sections of \citet{Verner1996} and solar abundances from \citet{Wilms2000} to account for the interstellar absorption, the hydrogen column density was fixed at 0.36 $\times 10^{-22}{\rm cm}^{-2}$ \citep{2013MNRAS.432.1144S}.
In Table~\ref{tab:table2}, we list all 45 observed X-ray bursts in the order of observation number and illustrate which bursts were found outside the GTIs.

% fit pre-burst spec
Given that a single power-law model component is satisfactory for fitting the pre-burst data, we employed this to investigate the spectral characteristics of the persistent emission.
We used the component $\sc cflux$ to calculate unabsorbed 2$-$10 keV, 10$-$35 keV, and 2$-$35 keV X-ray flux for each spectrum, as shown in Table~\ref{tab:table3}.

% fit burst spec in two ways
We used two methods to analyze the spectrum during the burst.
One is the standard method, commonly known as the `classical' approach.
The assumption of this method is that the persistent emission remains unchanged before and during the burst.
In this procedure, the pre-burst spectrum was utilized as the background spectrum in the fitting, and an absorbed single-temperature blackbody model, $\sc TBabs*bbodyrad$, for fitting the spectrum during the burst \citep{1986A&A...157L..10V,2002A&A...383L...5K}.
The above model includes the blackbody color temperature, ${kT}_{bb}$, and the normalization, ${K}_{bb}$, proportional to the square of the blackbody radius.
The bursts bolometric flux is calculated as
\begin{equation}
\label{eq1}
    F=1.076\times10^{-11}\left(\dfrac{kT_{\rm bb}}{1\,\rm{keV}}\right)^{4}K_{\rm bb}\quad \rm{ergs\;cm^{-2}\;s^{-1}},
\end{equation}
where ${K}_{\rm bb}= R_{\rm km}^{2} / D_{10}^{2}$, with $R_{\rm km}$ the effective radius of the emitter in km and $D_{10}$ is the distance to the source in units of 10 kpc (`classical' part of Table~\ref{tab:table4}).
The peak flux, the fluence, and the equivalent duration $\tau$ (the ratio of the burst fluence to the peak flux) in Table~\ref{tab:table2} are obtained by this method.

The other one is the `$f_{\rm a}$' method \citep{2013ApJ...772...94W}.
This method assumes that, while the persistent emission changes during the burst, the spectral shape remains unchanged and only the intensity changes.
In this approach, the background spectrum we utilized to fit the spectrum during the burst is the instrument background.
The fitting model, denoted as $\sc TBabs*(bbodyrad + f_{\rm a}*powerlaw)$, is utilized for the fitting process.
The parameters of the power law in this model were fixed as the results previously used to fit the 64-s pre-burst spectrum.
The variable dimensionless factor `$f_{\rm a}$' is used to represent the change of the intensity.
The fitting results of this method are also shown in Table~\ref{tab:table4}.

%-----------------------------------------------------
%%%-----RESULTS------%%%
\section{RESULTS}
\label{section:RESULTS}

%-----------------------------------------------------
\subsection{Light Curves}
\label{sec:Light Curves}

%introduce MAXI&BAT LC
\begin{figure*}
    \centering
	\includegraphics[width=\linewidth]{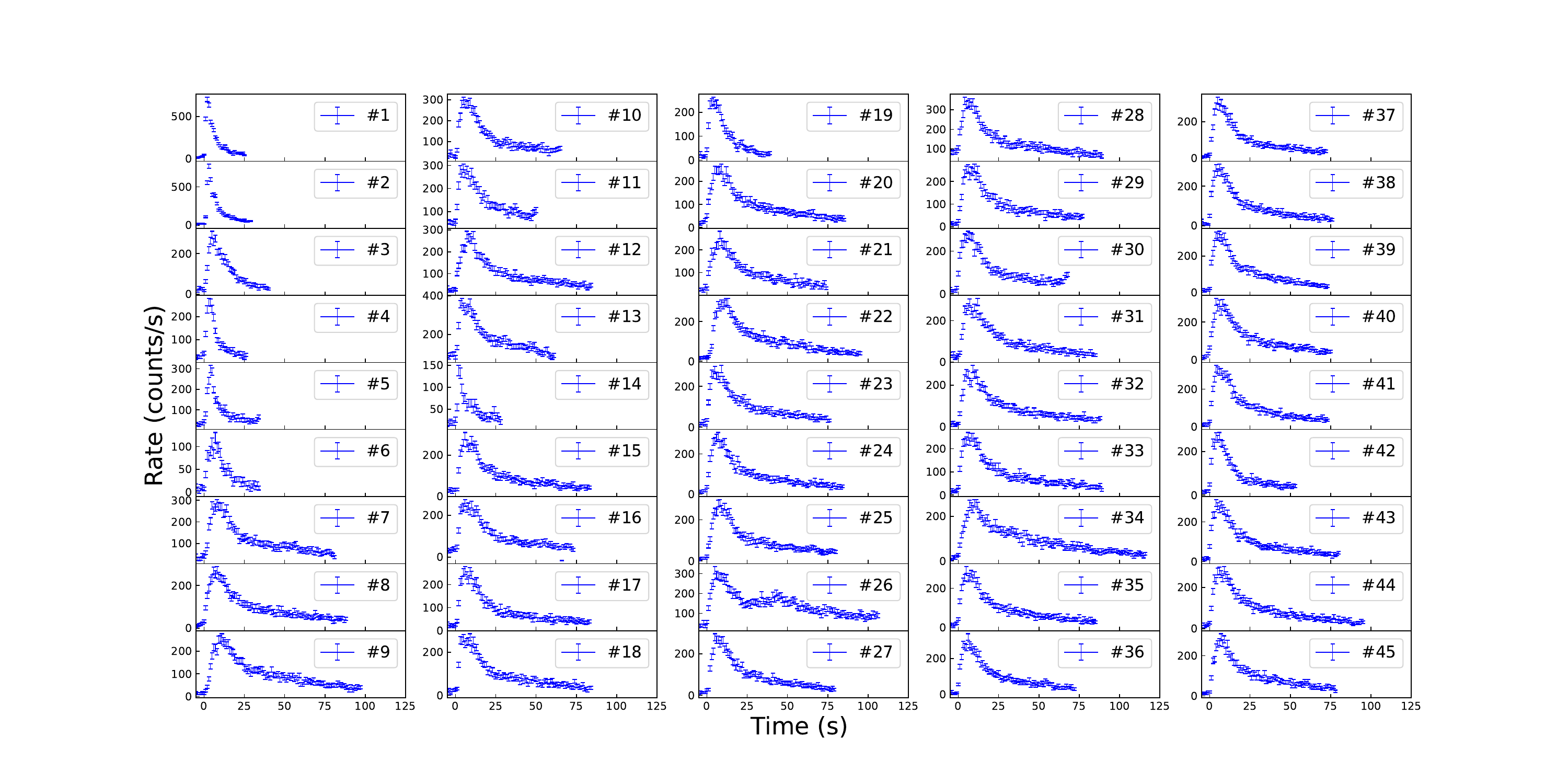} 
    \caption{Light curves of each X-ray burst from 4U 1636$-$ 53 observed with \textit{Insight}-HXMT sorted by time.
    These light curves are in the 2–10 keV energy band, binned at 1s time resolution, and expressed relative to $t_0$, the start time of each respective X-ray burst (see Table~\ref{tab:table2}).}
    \label{fig:figure3}
\end{figure*}

In Figure~\ref{fig:figure3}, we have utilized the available burst data to construct a comprehensive set of light curves.
These are 2$-$10 keV 1s light curves relative to their respective onset, sorted by their observation time.
In our sample, the majority of observed bursts had a duration exceeding 50 s, commonly referred to as frequent long bursts \citep{2008ApJS..179..360G}.
Compared to the frequent long bursts, we found that short bursts have diverse profiles.
The peak count rate of the first two bursts labeled \#1 and \#2 exhibit notably higher peak count rates, reaching 705 ct s$^{-1}$ and 730 ct s$^{-1}$, respectively, and their rise time are 1.25s and 1.75s.
Among these bursts, 36 bursts have a duration of more than 50 s, with rise times ranging from 3.0 s to 6.5 s.

%non-exponential part in \#26 
The typical light curve profile of the bursts has a fast rise and an exponential decay, but burst \#26 has a non-exponential decay in the 2-10 keV light curve, as shown in Figure~\ref{fig:figure3}.
To explore the origin of this non-exponential part, we examined it from two aspects, the light curve and the spectrum.
First of all, we generated the light curve of the same energy band and time bin across three detector boxes of the LE X-ray telescope.
We found that the non-exponential part of the light curves from the three boxes remained consistent, indicating that the signal of the non-exponential part comes from the same collimator line of sight.
\begin{figure}
    \centering
	\includegraphics[width=\columnwidth]{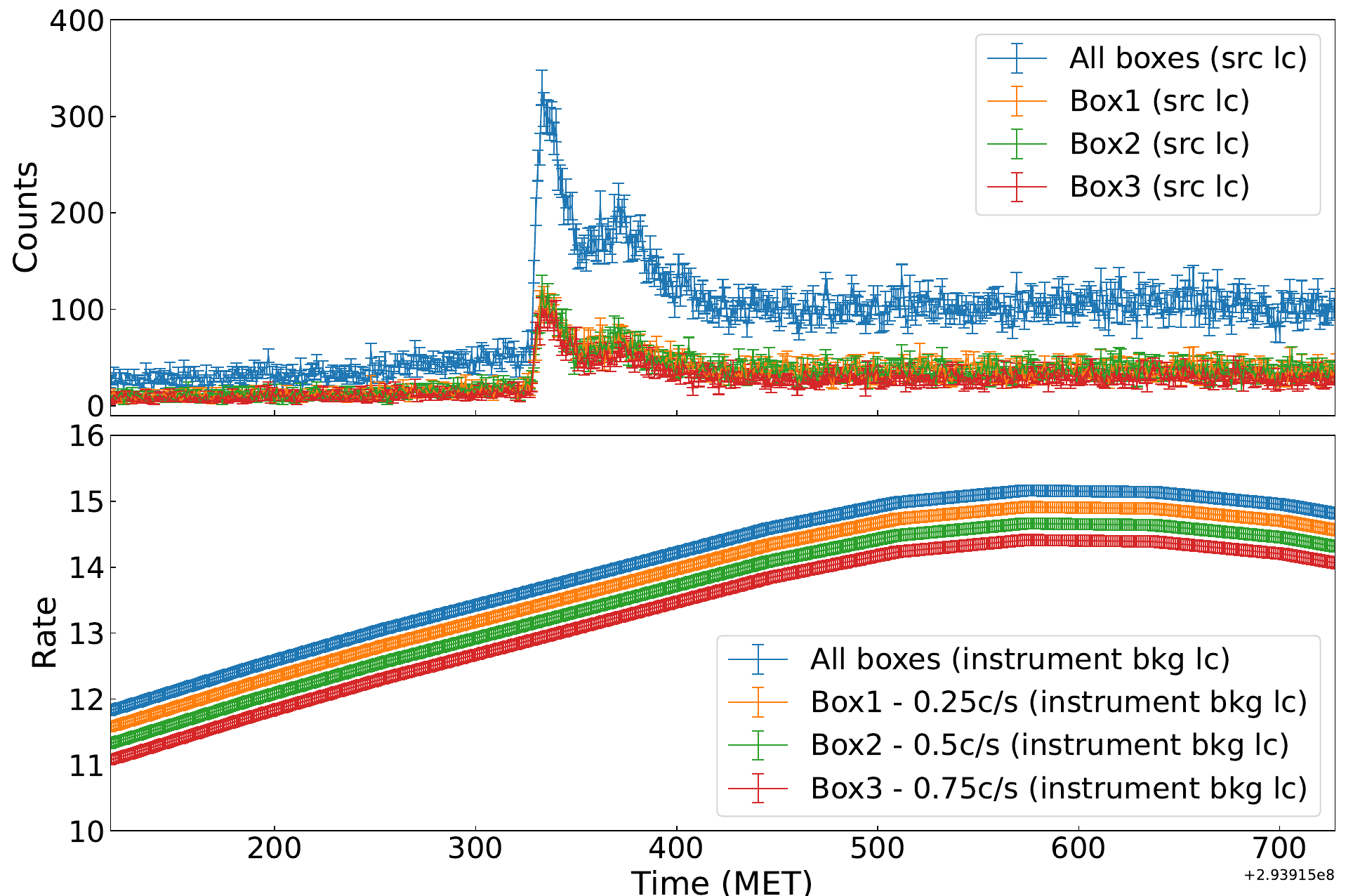}
    \caption{ Source light curves (src lc) and instrument background light curves (instrument bkg lc) of burst \#26.
    These light curves are in the 2–10 keV, binned at 1 s time resolution.
    The LE telescope contains three detector boxes placed at different angles.
    The blue, orange, green, and red solid lines represent the light curve of all boxes, box 1, box 2, and box 3, respectively.
    The upper panel shows the source light curves of different boxes.
    The lower panel shows the background light curves of different boxes.
    To better distinguish the light curves of different boxes, we slipped the curves by -0.25 ct s$^{-1}$, -0.50 ct s$^{-1}$ and -0.75 ct s$^{-1}$ from box 1, box 2 and box 3 as a whole, respectively.}
    \label{fig:figure4}
\end{figure}
We also investigated the instrument background light curves of LE and ME, which are both smooth, but show an overall upward trend, as shown in Figure~\ref{fig:figure4}.
Burst \#26 presented a unique challenge due to its occurrence outside the GTI, making it difficult to accurately assess its instrument background spectrum.
The following analysis does not include the decay phase of this burst.

%% 2,  figure 4 (similar profile bursts)
\begin{figure} 
    \centering
	\includegraphics[width=\columnwidth]{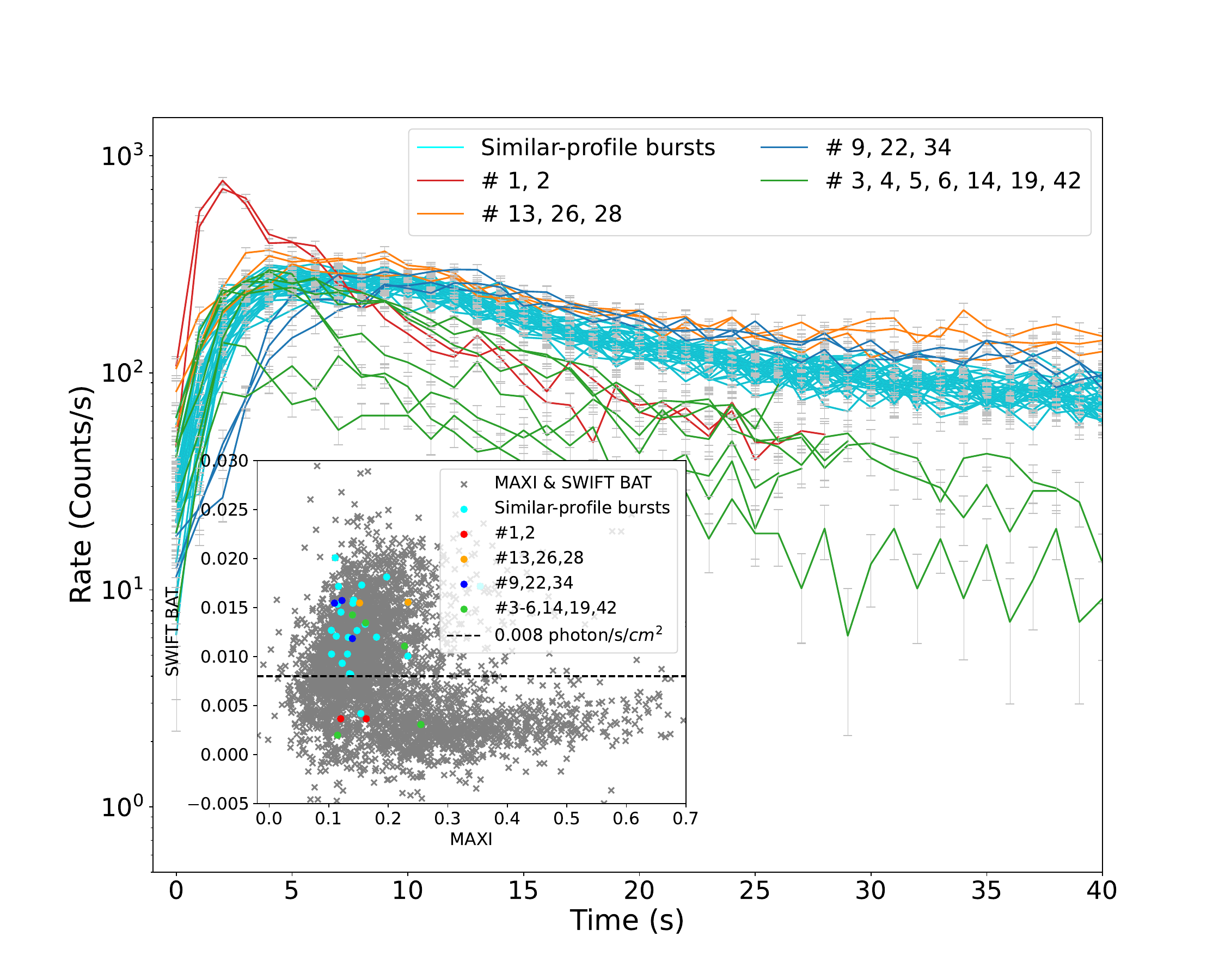}
    \caption{X-ray burst light curves in the 2$-$10 keV aligned by their start time and with the instrument background count rate subtracted. 
    The cyan color represents 30 bursts with similar profiles. 
    The red color represents 2 PRE bursts.
    The orange color represents 3 bursts with a long duration and higher rate in the decay phase.
    The blue color represents 3 bursts with a long rise time of about 6 s.
    The green color represents 7 bursts with short duration.
    The intensity-intensity diagram in the figure marks the positions of these bursts in the 2$-$20 keV \textit{MAXI} and 15$-$50 keV BAT 1-day intensity plot.
    The black dashed line represents the rate of BAT at 0.008 photon s$^{-1}$cm$^{2}$.}
    \label{fig:figure5}
\end{figure}
In 4U 1636$-$53, we observed a variety of light curve profiles for Type-I bursts. 
To compare and study the different burst light curve shapes, we aligned the start times of all bursts and displayed them in Figure~\ref{fig:figure5}.
These bursts can be roughly divided into three categories: similar-profile bursts, PRE bursts (see Section \ref{sec:PRE bursts}), and other non-PRE bursts.
The 30 cyan burst light curves in the figure have a similar profile, characterized by a rising time of approximately 4 s, followed by a decay phase lasting over 35 s.
Bursts \#1 and \#2 are PRE bursts, which have a quick rise time and higher peak count rates.
All other bursts can be categorized as the other non-PRE bursts.
Bursts \#3-6, \#14, \#19, and \#42 have short duration.
Bursts \#13, \#26, and \#28 have about 300 ct s$^{-1}$ peak count rate, a duration of more than 35 s, and a higher count rate in the decay phase than similar-profile bursts.
Bursts \#9, \#22, and \#34 have a long rise of about 6 s and a higher decay phase than similar-profile bursts.

If we tentatively consider the intensity of \textit{Swift}-BAT in Figure~\ref{fig:figure2} and Figure~\ref{fig:figure5}, using 0.008 ct s$^{-1}{\rm cm}^{-2}$ as a threshold (black dashed line in Figure~\ref{fig:figure5}), bursts above this line happen when the source in the hard state, while those below when the source in the soft state.
Two PRE bursts and short bursts were observed in the soft state.
The frequent long bursts were in the hard state, including similar-profile bursts.

%-----------------------------------------------------
\subsection{Burst decay phase}
\label{sec:Burst decay phase}

\begin{figure} 
    \centering
	\includegraphics[width=\columnwidth]{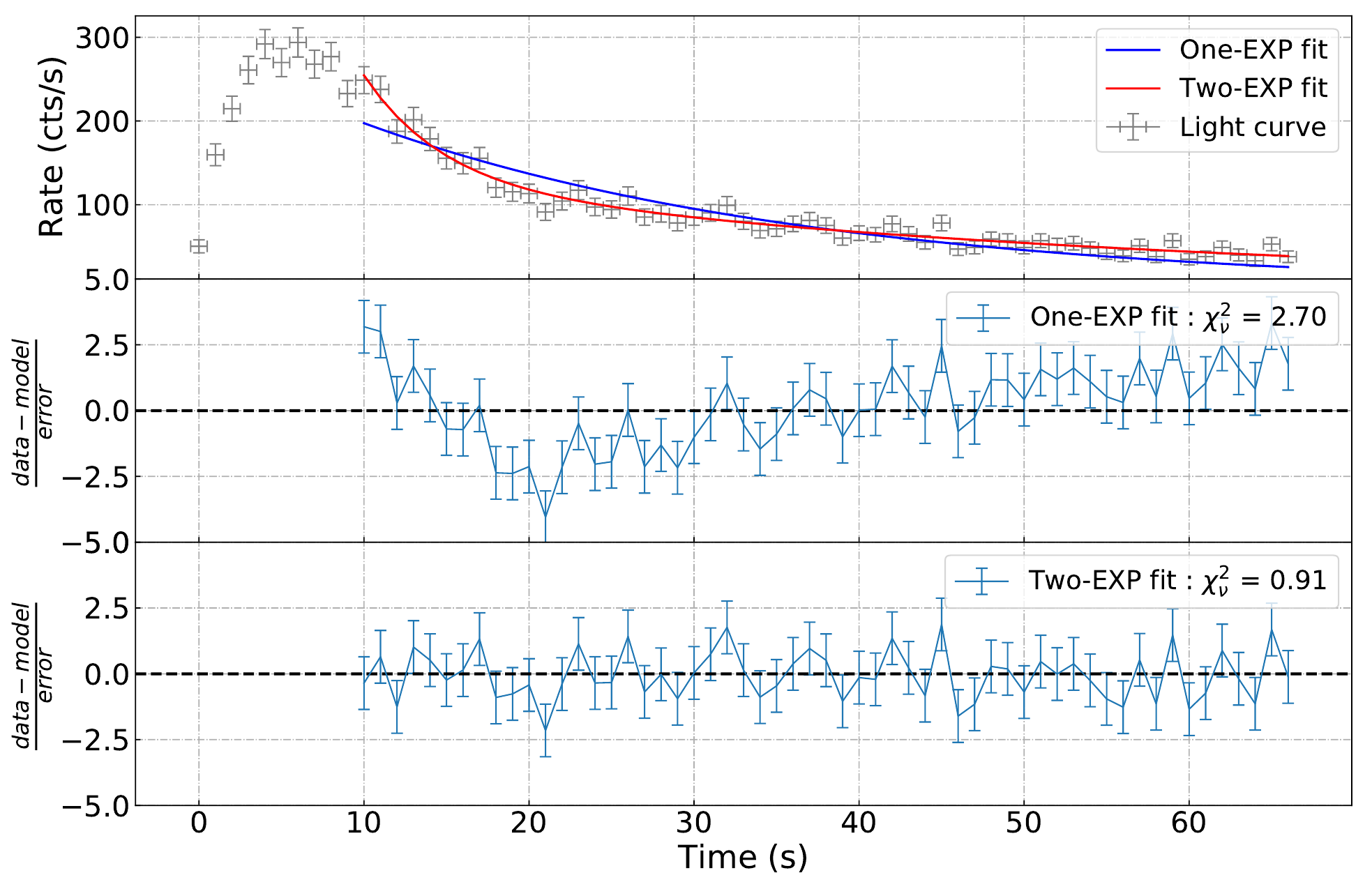}
    \caption{Example of a fit to the decay phase of a light curve of burst \#38.
    The top panel shows the light curve of the burst (grey point) and the best-fit models for a single exponential component (blue curve; Eq. (\ref{eq2})) and the two-exponential component (red curve; Eq. (\ref{eq3})).
    The second and third panels show the residuals of the two models.
    }
    \label{fig:figure6}
\end{figure}
After aligning the light curve in Figure~\ref{fig:figure3}, we observed that the decay phase of the bursts exhibits a visually discernible two-stage pattern.
Our investigation aims to validate the existence of these two decay phases and quantify the duration of the initial stage, which corresponds to the decay of the turning point.
The approach that we utilized consists of the following steps.

First, we determine the start time and the duration of the decay phase.
We identified the last data point above 90\% peak rate as the start time point of the decay phase.
The duration of the total decay phase is from this point to the time that drops below the pre-burst mean rate.

Second, we fit the decay phase.
Initially, we used the exponential component of the BURS model in QDP \footnote{https://heasarc.gsfc.nasa.gov/ftools/others/qdp/qdp.html}, which represents a model to describe the burst profile.
The function of the One-EXP (one exponential) model is
\begin{equation}
\label{eq2}
       F_1(t)=BN*e^{- \frac{t-{T_0}}{DT}},
\end{equation}
with normalization BN, initial time of the decay ${T_0}$, and exponential decay time DT.
Subsequently, we calculated $\chi_v^2$ (the reduced chi-square); if $\chi_v^2$ \textgreater 2, we judged that this single exponential model was inadequate for fitting the whole decay phase of the burst.
Consequently, we added another exponential component into the initial model to describe possible two decay phases.
The function of the Two-EXP (two exponential) model is
\begin{equation}
\label{eq3}
      F_2(t)={BN_1}*e^{- \frac{t-{T_0}}{DT_1}} + {BN_2}*e^{- \frac{t-({T_0}+{DT_1})}{DT_2}},
\end{equation}
with normalization ${BN_1}$, ${BN_2}$, exponential decay time ${DT_1}$ (initial duration), ${DT_2}$ and secondary decay onset time $({T_0}+{DT_1})$.
The former exponential component represents the first decay part, the latter one accounts for the rest, and $({T_0}+{DT_1})$ also is the time of the turning point between the two phases.

 \begin{figure} 
	\includegraphics[width=\columnwidth]{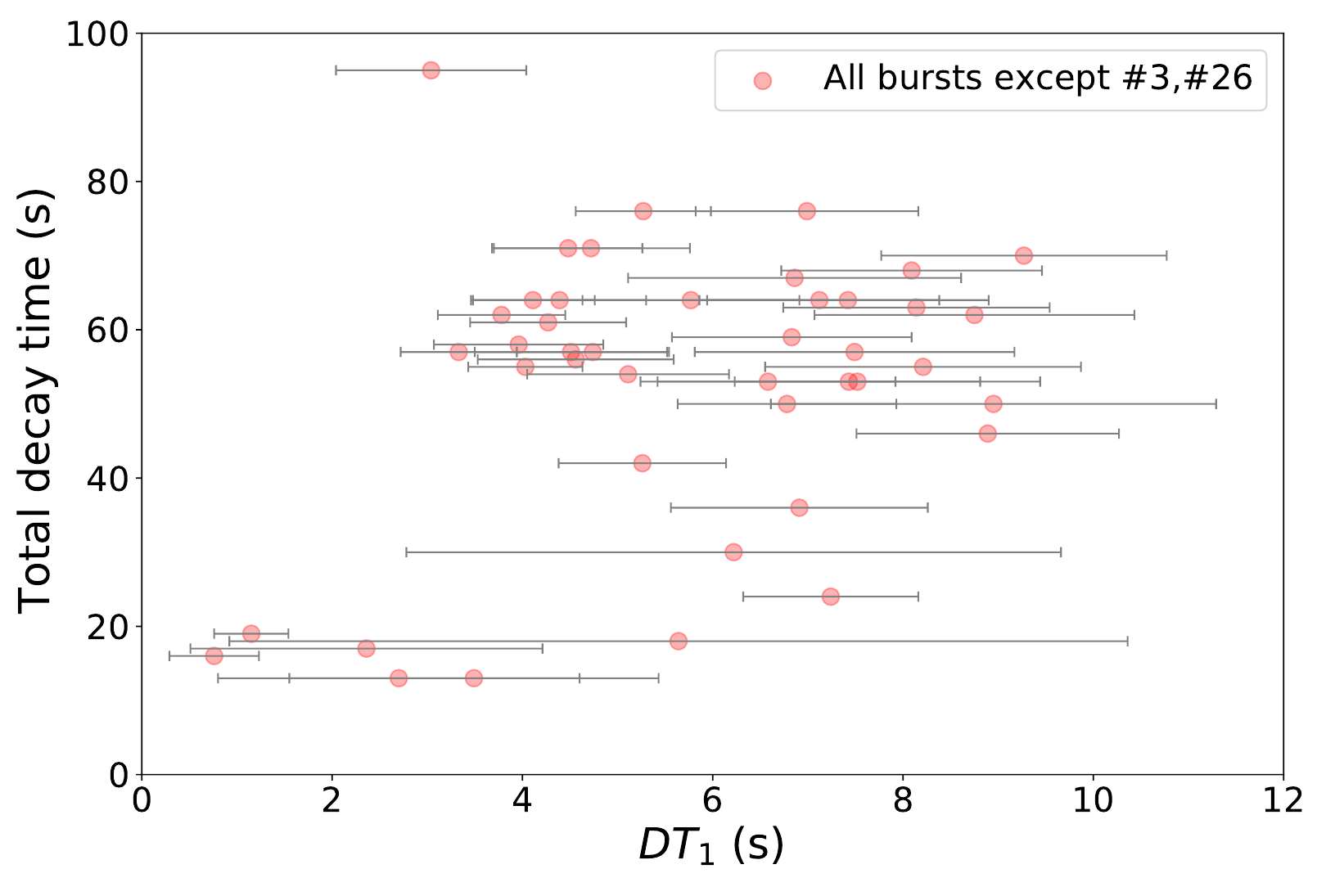}
    \caption{The correlation between the total decay time and $DT_1$ for 43 bursts of 4U 1636$-$53.
    This plot dose not include bursts \#3 and \#26.
    Burst \#3 can be fit well by the one exponential model, without a second exponential component (for burst \#26 see Section \ref{sec:Light Curves}).
    }
    \label{fig:figure7}
\end{figure}

After fitting the decay phase of all bursts with the One-EXP model, most bursts' $\chi_v^2$ are greater than 2, except bursts \#3, \#6, \#9, \#14, and \#34.
To investigate whether all bursts have a double decay phase, we tried to fit all bursts with the Two-EXP model.
The fitting example is shown in Figure~\ref{fig:figure6}.
The two-exponential component model fits the decay phase well.
In all the bursts, the $\chi_v^2$ distribution of the One-EXP model is centered at around 2.5, whereas the Two-EXP model fits better, and the values of $\chi_v^2$ centered around 1.
We found that the distribution of $DT_1$ is consistent with a Gaussian distribution centered around 6 s.

We plotted the first exponential decay time $DT_1$ as a function of the total decay time, as shown in Figure~\ref{fig:figure7}.
Notably, burst \#3, unlike the others, can be effectively fitted with the One-EXP model alone, without requiring a second exponential component.
This plot contains the results of 43 bursts except for bursts \#3 and \#26.
We calculated the correlation using the Pearson correlation coefficient \citep{23fbf67b-cef6-3e43-bbf3-1b85519a1108}.
We found an unrelated correlation between the total decay time and $DT_1$, and the correlation coefficient is 0.323, with a chance probability of 0.035.
When the total decay time is below 20 s, $DT_1$ is around 2 s. 
As the total decay time increases to 80 s, $DT_1$ is around 6 s. 
We noticed that $DT_1$ of a burst exceeding 80 s is $\sim$ 4 s.

%-----------------------------------------------------
\subsection{The relation between pre-burst spectrum and burst profile}
\label{sec:The relation between pre-burst spectrum and burst profile}

 \begin{figure} 
	\includegraphics[width=\columnwidth]{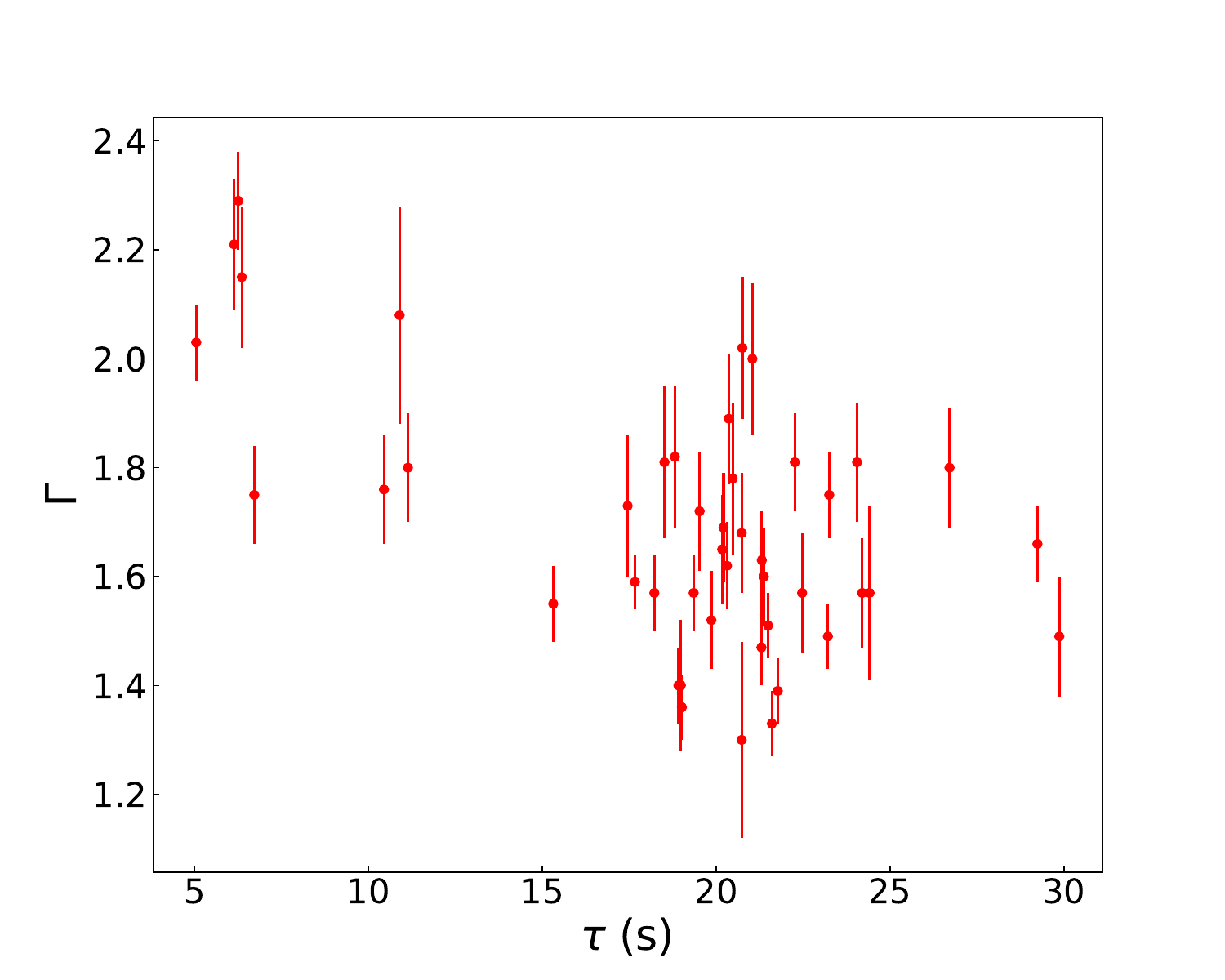}
    \caption{Correlations between the power-law photon index $\Gamma$ of the pre-burst spectrum and the equivalent duration $\tau$ for 45 bursts of 4U 1636$-$53.
    }
    \label{fig:figure8}
\end{figure}

To investigate if the pre-burst emission affects the burst profile, we calculated the correlation coefficient between the power-law photon index ($\Gamma$) of the pre-burst spectrum and the equivalent duration (the ratio of the burst fluence to the peak flux, $\tau$) of the bursts. 
Figure~\ref{fig:figure8} shows $\tau$ as a function of $\Gamma$. 
The correlation coefficient is -0.57, with a chance probability of 3.73 $\times10^{-5}$, which shows there is an anti-correlation.
The pre-burst spectrum is relatively flat, as this source is in the hard state.
As $\Gamma$ decreases from 2.2 to 1.5, the source gradually changed from the soft state to the hard state, and the equivalent duration of the burst continued to increase.
In the same source, \citet{2015MNRAS.454..541L} found that as the accretion rate gradually decreased, the source gradually evolved from the soft state to the hard state, and the equivalent duration of X-ray bursts gradually increased from 2 to 25.
Their findings align with ours, indicating that short bursts tend to appear in the soft state, while frequent long bursts are associated with the hard state.

%-----------------------------------------------------
\subsection{PRE bursts}
\label{sec:PRE bursts} 

\begin{figure*}
    \centering
    \begin{minipage}{0.49\linewidth}
     	\includegraphics[width=0.9\columnwidth]{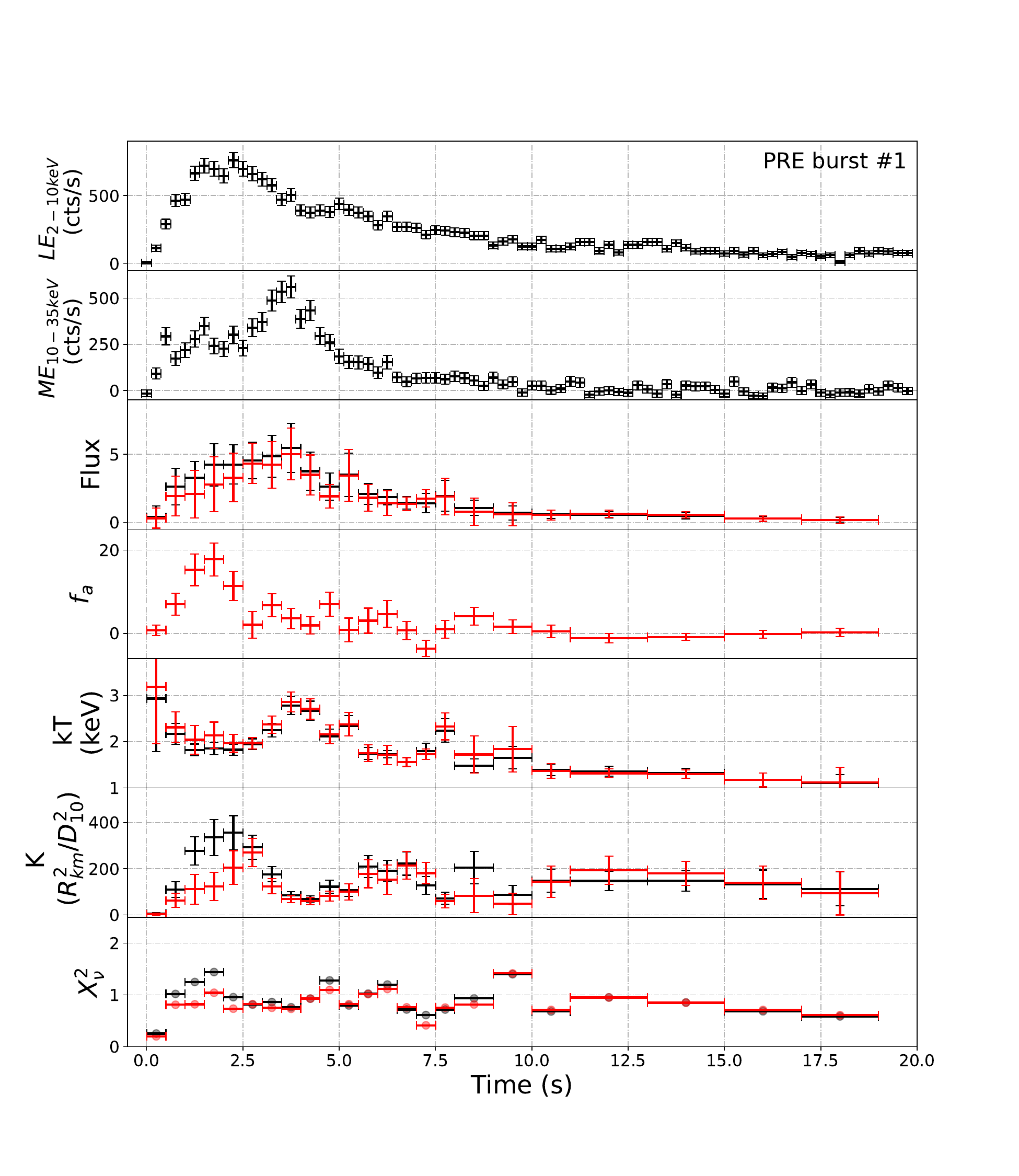}
    \end{minipage}
    \quad
    \begin{minipage}{0.49\linewidth}
	\includegraphics[width=0.9\columnwidth]{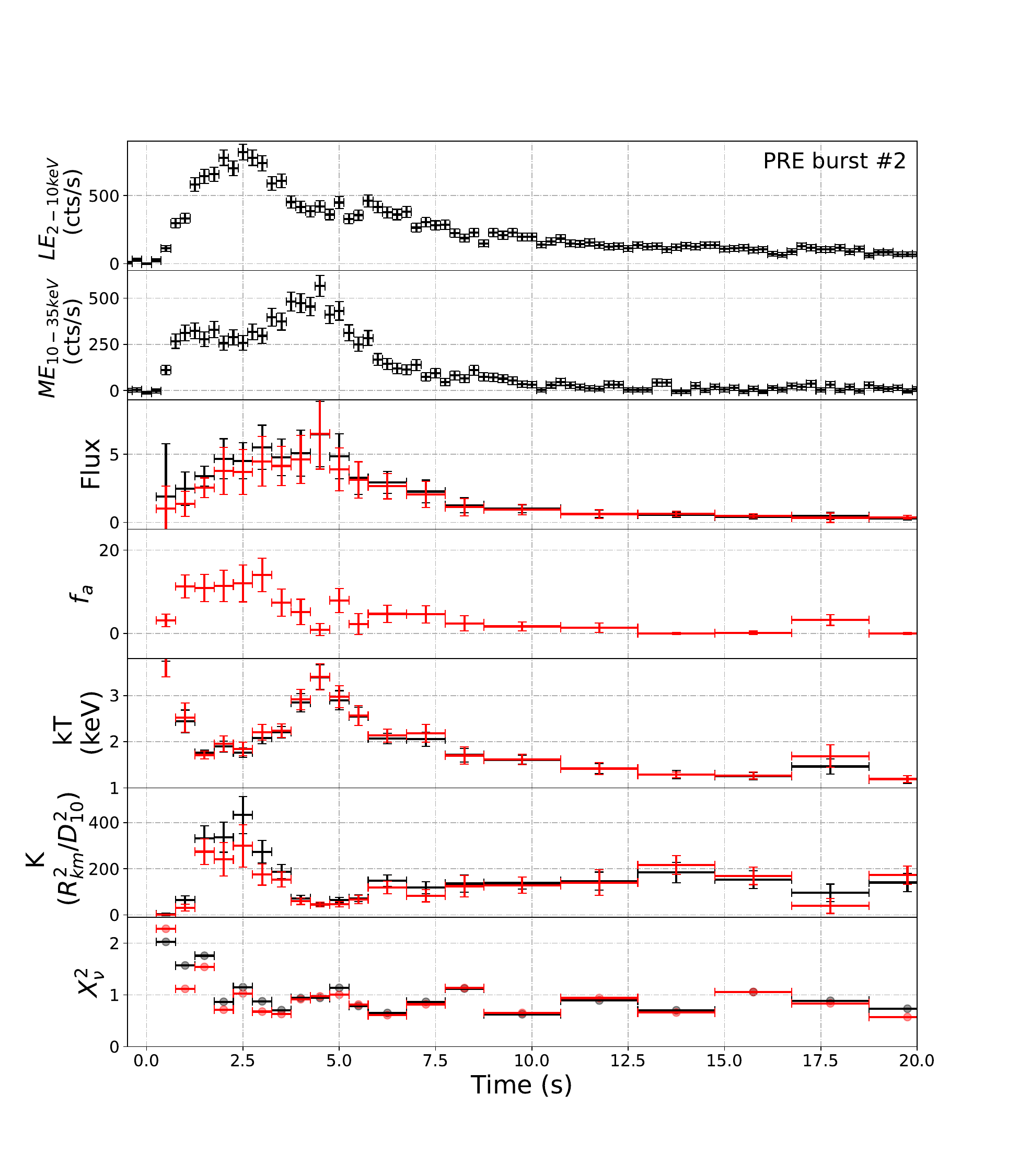}
    \end{minipage}
\caption{Comparison of the results of two methods for fitting the two PRE bursts in 4U 1636$-$53, the left and right figures are burst \#1 and \#2, respectively.
From top to bottom, each panel in the figure are LE (2$-$10 keV) light curve at 0.25s resolution, ME (10$-$35 keV) light curve at 0.25s resolution, bolometric flux ($\times10^{-8}$erg s$^{-1}$cm$^{-2}$), the value of $f_{\rm a}$, blackbody temperature, normalization (${K}_{\rm bb}$) of the blackbody, and reduced chi-square.
The black and red points with error bars in panels 3, 4, and 5 are the fitting results of the `classical' method and the `$f_{\rm a}$' method respectively.
}
\label{Figx:figure9}
\end{figure*}

%light curve 
The first two panels of Figure~\ref{Figx:figure9} show the light curves of bursts \#1 and \#2.
The LE light curve of these bursts rises fast, within 2 s.
Meanwhile, the ME light curves exhibit a `dip' before reaching the peak.
Similar double peak light curve phenomena have been found in MXB 1743$-$29, MXB 1728$-$34, MXB 1850$-$08 \citep{1980ApJ...240L..27H}, 4U 1820$-$30 \citep{1987ApJ...314..266H}, SAX J1808.4$-$3658 \citep{2019ApJ...885L...1B}, and 4U 1608$-$52 \citep{2022ApJ...936...46C}.
The limited energy coverage during the observation of the PRE burst may provide a possible explanation for this double peak in the ME light curves. 
The radiation energy distribution of the burst shifts towards the lower energy band during the expansion phase of the PRE burst, resulting in a decrease in flux as detected by the ME (10$-$35 keV) \citep{2003A&A...399..663K}.
When the PRE burst’s luminosity exceeds the Eddington limit, the radiation from the burst could drive the inner disk material around the NS to form an outward moving wind with a large column density \citep{2005ApJ...626..364B}. 
The radiation may also enhance the emission of comptonization through the reprocessing of burst emission in the accretion environment \citep{2018ApJ...856L..37K}.
We strategically chose varying exposure times to extract spectra, to study the two possible PRE bursts in more detail.
Specifically, we opted for exposure times of 0.5s, 1s, and 2s, aligning with our prior methodology in Section \ref{sec:OBSERVATION AND DATA REDUCTION}.
After fitting with the `classical' method, burst \#1 had a peak at a bolometric flux of 5.36$\pm$1.22 $\times10^{-8}$erg s$^{-1}$cm$^{-2}$, and burst \#2 5.85$\pm$1.41 $\times10^{-8}$erg s$^{-1}$cm$^{-2}$.

% describe LC and paras after onset
The last five panels of the two plots of Figure~\ref{Figx:figure9} are the parameters of the two fitting methods, with the black and red dots representing, respectively, the `classical' method and the `$f_{\rm a}$' method.
The peak of the blackbody temperature corresponds to the peak value of the ME light curve.
From the onset of the burst the blackbody temperature  rises to a peak and subsequently declines to the pre-burst level.
The blackbody normalization mirrored this trend in reverse.
Such dynamic variations are consistent with the characteristic behavior of the PRE burst \citep{1984ApJ...281..337B}.

%-----------------------------------------------------
\subsection{Time Resolved Spectroscopy}
\label{sec:Time Resolved Spectroscopy} 

 \begin{figure} 
	\includegraphics[width=\columnwidth]{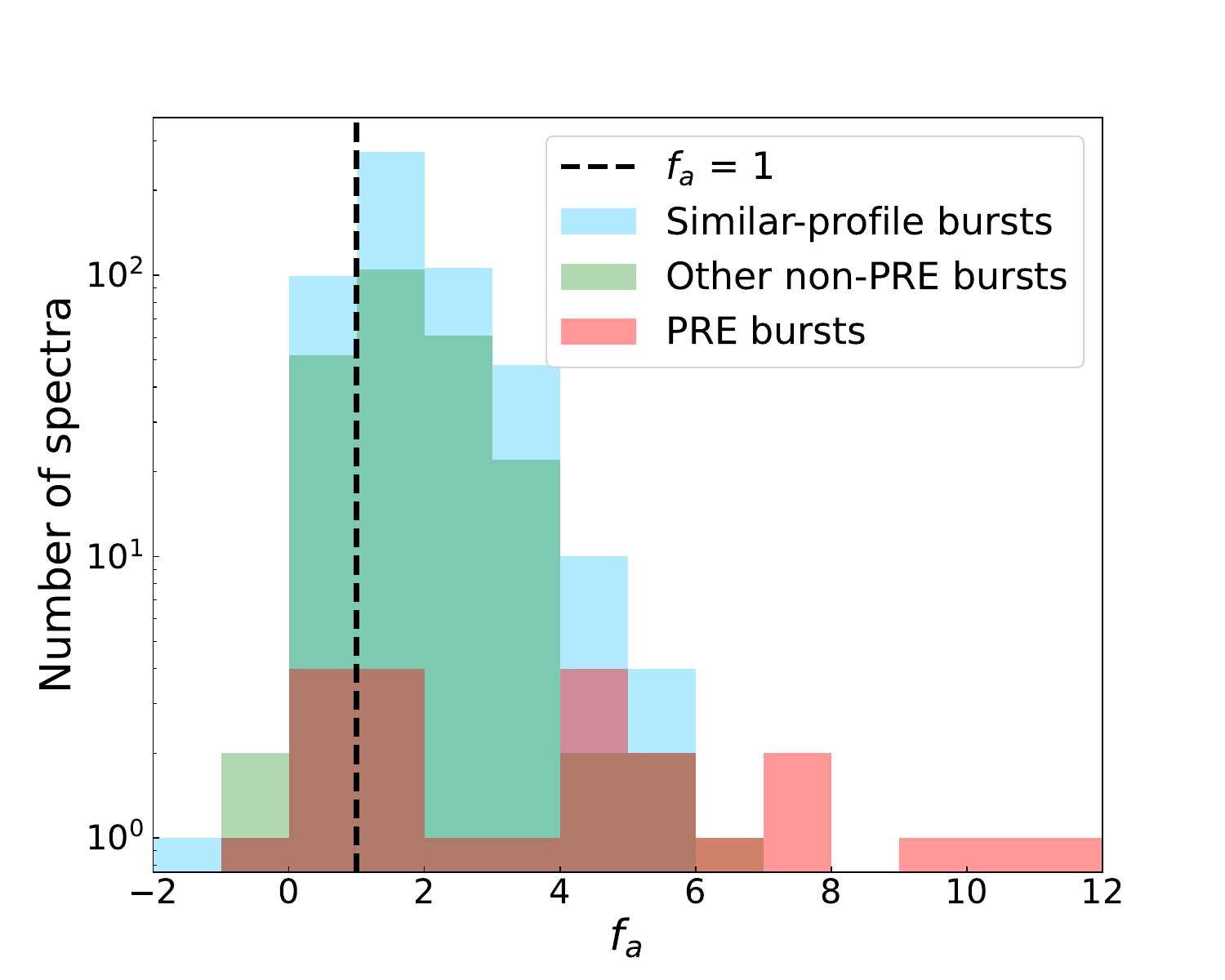}
    \caption{The histogram of the inferred $f_{\rm a}$ values obtained using the `$f_{\rm a}$' method.
Blue, green, and red represent, respectively, the fitting results of 30 similar-profile bursts, other non-PRE profile bursts, and the 2 PRE bursts.
    }
    \label{fig:figure10}
\end{figure}

%`classical' and 'fa' fit method
After fitting the spectra during the burst with the `$f_{\rm a}$' method, we found that the $f_{\rm a}$ values are mostly greater than 1.
Figure~\ref{fig:figure10} shows the distribution of $f_{\rm a}$ according to the burst classification.
Similar-profile and other non-PRE types of bursts have alike $f_{\rm a}$ distributions. 
The $f_{\rm a}$ value of PRE bursts ranges from $- 1$ to 12. 
Some $f_{\rm a}$ values in this distribution are negative.
After examining the corresponding spectra, we found that these spectra occur at the initial and final stages of the burst, exhibiting a low count rate.
Therefore, these $f_{\rm a}$ values are deemed unreliable.

%persistent flux and max fa
\begin{figure*}
    \centering
    \begin{minipage}{0.49\linewidth}
     	\includegraphics[width=0.9\columnwidth]{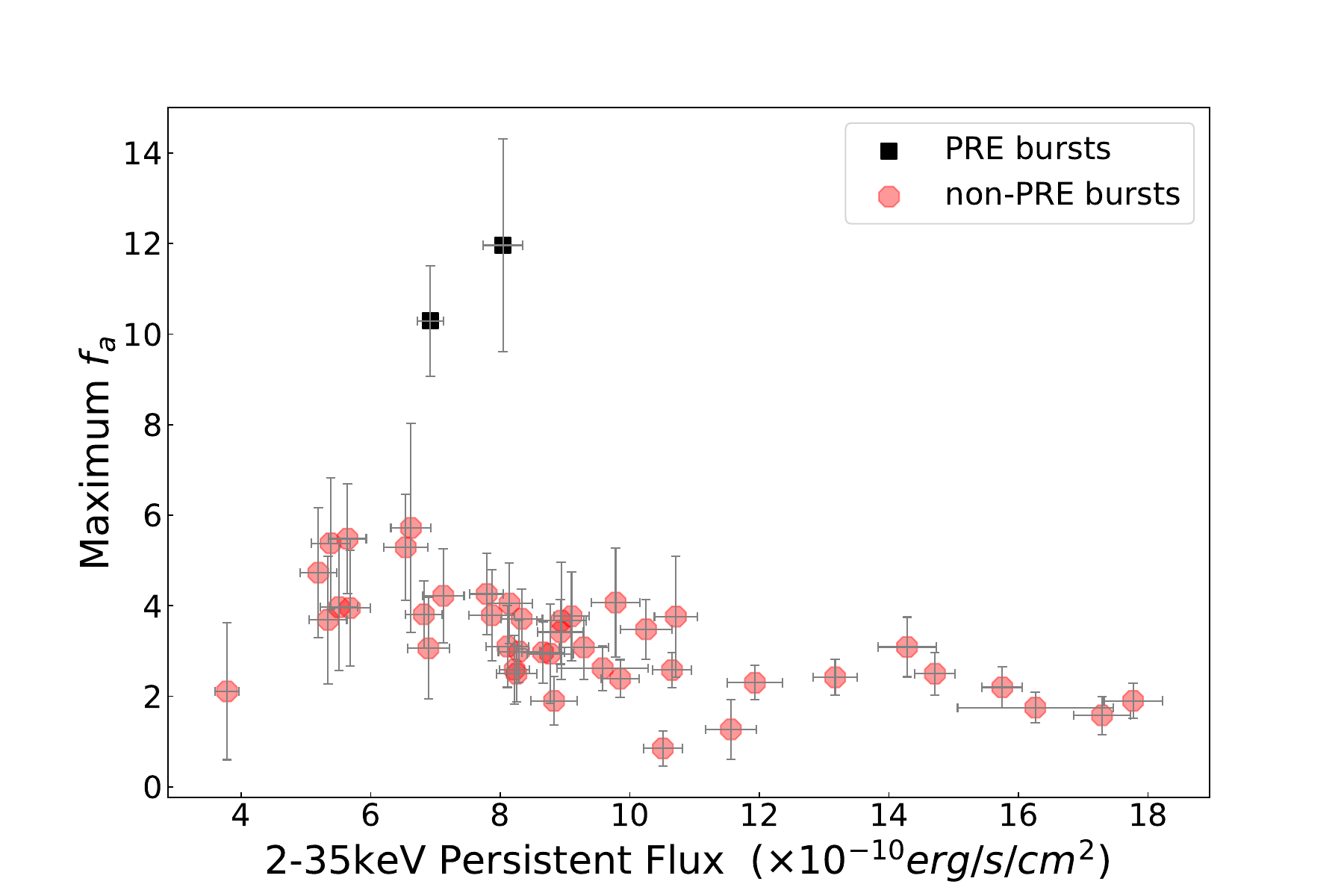}
    \end{minipage}
    \quad
    \begin{minipage}{0.49\linewidth}
	\includegraphics[width=0.9\columnwidth]{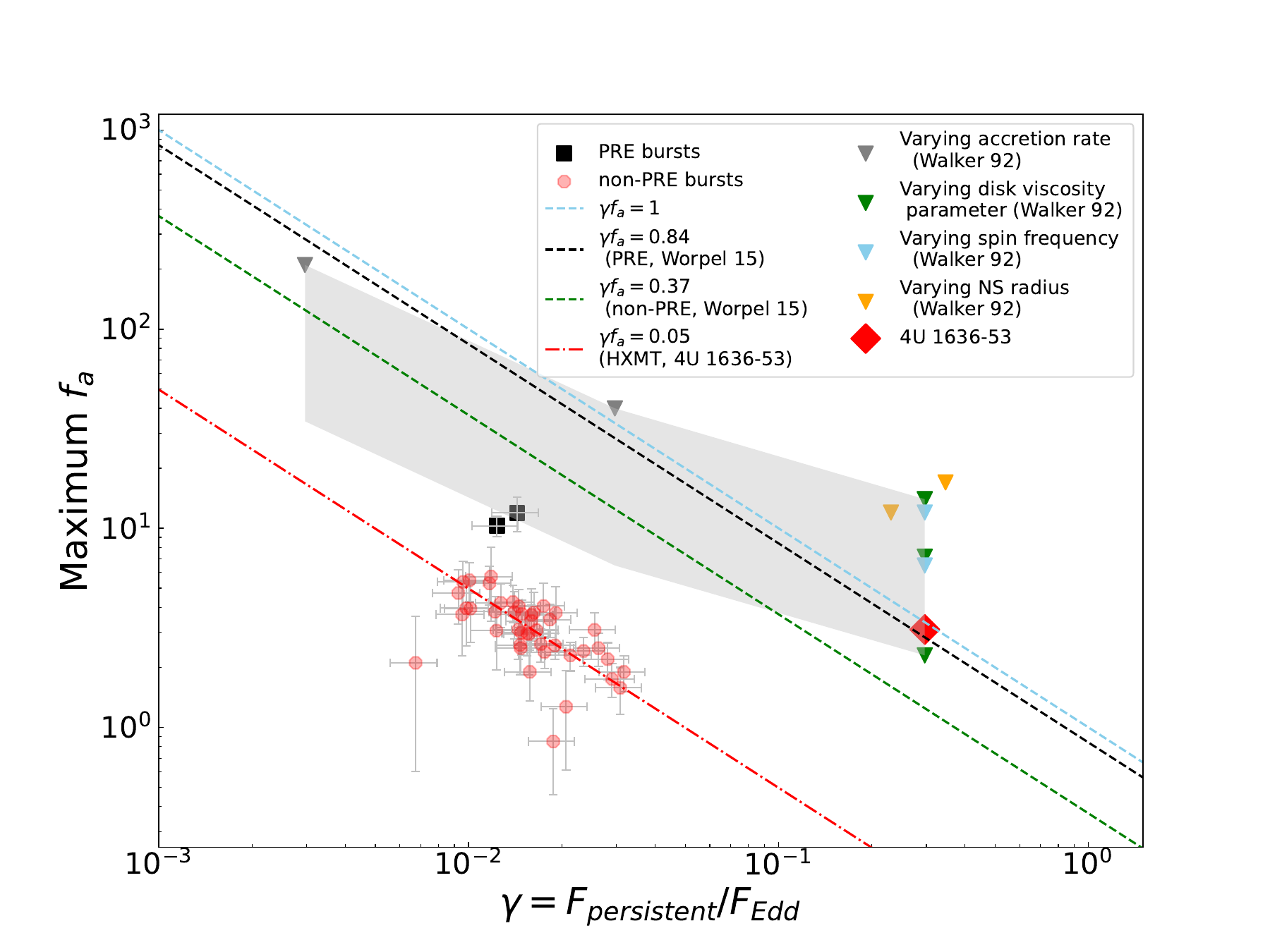}
    \end{minipage}
\caption{The left plot shows the maximum $f_{\rm a}$ against persistent flux.
Black squares and red dots represent, respectively, the PRE bursts and non-PRE bursts.
The right plot is the maximum $f_{\rm a}$ vs $\gamma = F_{\rm persistent} / F_{\rm Edd}$ (the pre-burst accretion rate as a fraction of $\dot M_{\rm Edd}$), $F_{\rm Edd}$ is the mean value of the maximum flux obtained by the classical method of two PRE bursts.
Among them, gray, green, skyblue, and orange inverted triangles represent the simulated values at varying accretion rate, varying disk viscosity, varying spin frequency, and varying NS radius respectively \citep{1992ApJ...385..642W}.
The red diamond point using theoretical simulated values (the spin frequency $=$ 1.5 ms, the NS radius $r_{*} = 7$ km and the dimensionless units of accretion rate $\dot m = 1$) in Table 1 of \citet{1992ApJ...385..642W} that are close to the system parameters (the spin frequency $\sim$ 1.7 ms) in 4U 1636$-$53.
The skyblue, black, and green dotted lines represent $\gamma f_{\rm a}$ equal to 1, 0.85 (upper limit of PRE bursts), and 0.37 (upper limit of non-PRE bursts), respectively \citep{2015ApJ...801...60W}.
The red dotted line represents the average value $\gamma f_{\rm a} = 0.05$ of non-PRE bursts.
}
\label{Figx:figure11}
\end{figure*}

In the left of Figure~\ref{Figx:figure11}, we draw the relation between the persistent emission flux and the maximum $f_{\rm a}$ value.
The black square and red point in these plots represent PRE bursts, non-PRE state bursts.
We only calculated the correlation coefficient of all non-PRE bursts.
The correlation coefficient between the persistent emission flux and the maximum $f_{\rm a}$ is -0.62, with a chance probability of 8.30 $\times 10^{-6}$.
This indicates a negative correlation, implying that the strength of the burst's impact increases as the persistent emission weakens. 

We followed \citet{2008ApJS..179..360G} to calculate the dimensionless persistent flux $\gamma = {F}_{\rm persistent} / {F}_{\rm Edd}$, which can be approximately equal to the accretion rate as a fraction of the Eddington rate, $ \dot{M}/ \dot{M}_{\rm Edd}$.
${F}_{\rm Edd}$ is the Eddington flux from the results of PRE bursts fitting by the ‘classical’ method.
In the right figure of Figure~\ref{Figx:figure11}, we plot the relation between the maximum $f_{\rm a}$ and $\gamma$.
The skyblue, black, and green dotted lines represent $\gamma f_{\rm a}$ equal to 1, 0.85 (upper limit of PRE bursts), and 0.37 (upper limit of non-PRE bursts), respectively \citep{2015ApJ...801...60W}.
% observation
In \textit{Insight}-HXMT’s results, the PRE bursts are above the non-PRE bursts.
The results of \textit{Insight}-HXMT are below the upper limit of PRE and non-PRE in \citet{2015ApJ...801...60W} results.
% theoretical
The theoretical results of \citet{1992ApJ...385..642W}, on the assumption that the $f_{\rm a}$ is completely caused by accretion enhancement, which corresponds to $\Delta \dot{M}_*/ \dot{M}_*$ (the ratio of peak to quiescent mass transfer rates).
For calculating $\gamma$, the NS mass $M_{*}$ is set to 1.4 $M_{\sun}$ \citep{1992ApJ...385..642W}.
The accretion rate in dimensionless units is $\dot{m} = {\dot{M}c^2}/{L_{\rm Edd}}$.
Since the mass ${M}_*$ and the radius ${R}_*$ were given, $\gamma = {GM_{*}}\dot{m}/{c^2R_{*}}$ \citep{2013ApJ...772...94W}.
The simulated value in Table 1 of theoretical results to calculate the simulated $\gamma$ under different conditions.
The NS radius $r_*$ were 6 km, 7 km and 9 km, the accretion rate $\dot{m}$ were 1, ${10}^{-1}$ and ${10}^{-2}$, the disk viscosity $\beta$ were 1, ${10}^{-2}$ and ${10}^{-6}$ and the spin frequency were 1.5 ms, 3 ms and ${10}^{5}$ ms.
So, we used the gray area and green, skyblue, and orange inverted triangle to draw the results of \citet{1992ApJ...385..642W}.
We used $\sim$ 1.7 ms spin period of 4U 1636$-$53 \citep{1997IAUC.6541....1Z,1998ApJ...498L.135S,2002ApJ...577..337S}, and marked the point close to 4U 1636$-$53 in the theoretical prediction with the red diamond.
Looking at the overall distribution of these bursts first, only PRE bursts are near the theoretical prediction range, and non-PRE bursts are below the range.
However, the anti-correlation of \textit{Insight}-HXMT’s results is consistent with the theoretical prediction.

%-----------------------------------------------------
%%%-----DISCUSSION------%%%
\section{DISCUSSION AND CONCLUSIONS}
\label{sec:DISCUSSION}

We analyzed 45 bursts observed by \textit{Insight}-HXMT in the LMXB 4U 1636$-$53. 
We found 30 bursts with long duration exhibit a similar profile, these bursts mainly occur in the hard state. 
The decay phase of the frequent long bursts shows a two-stage behavior. 
For the first time, we observed the duration of the initial stage is approximately 6 s. 
The majority of `$f_{\rm a}$’ values exceed 1. 
The non-PRE bursts show an inverse correlation between the maximum `$f_{\rm a}$’ and the persistent emission flux.
The enhancement of persistent emission may be mainly attributed to the inner corona zone.

\subsection{Burst properties at different accretion states}
\label{sec:Burst properties at different accretion states}

We observed 70 bursts during the 1050 ks observation with \textit{Insight}-HXMT. 
The recurrence time was 4.17 hr and the burst rate was 0.24 hr$^{-1}$. 
This value closely matches the 0.26 hr$^{-1}$ in 4U 1636$-$53 reported by \citet{2021ASSL..461..209G}. 
We categorized the quasi-simultaneous observations of \textit{Insight}-HXMT in the Intensity-Intensity diagram into the hard and the soft states, using BAT's 0.008 counts s$^{-1}$cm$^{-2}$ as a reference. 
In the soft state, insight-HXMT/LE had an exposure time of 100.92 ks, observed 6 bursts, with a recurrence time of 4.67 hr and a burst rate of 0.21 hr$^{-1}$. 
In the hard state, the exposure time for insight-HXMT/LE was 453.84 ks, with 33 observed bursts, a recurrence time of 3.82 hr, and a burst rate of 0.26 hr$^{-1}$. 
This shows that bursts occur more frequently in the hard state.
This phenomenon may be attributed to the increase in accretion rate, which leads to a larger area of stable burning and a decrease in unstable burning in the soft state. 
Consequently, this decrease in unstable burning results in a reduced burst rate \citep{2018ApJ...857L..24G}.

Given that there are more bursts in the hard state, we utilized the recurrence time of these bursts as a rough approximation for determining the $\alpha$-value (the ratio of the integrated persistent flux to the burst fluence), the H fraction X, and the ignition depth y.
Within our sample, the mean persistent emission flux is 11 $\times10^{-10}$erg s$^{-1}$cm$^{-2}$, with a mean fluence of 46 $\times10^{-8}$erg cm$^{-2}$. 
Our calculations are based on the assumption that the mass of the NS is 1.4 $M_{\sun}$, the radius is 10 km, the distance is 6 kpc, and (1+z) is 1.31. 
Based on eqs. 4-6 and a mean bolometric correction of 1.38 in \citet{2008ApJS..179..360G}, the calculated values for $\alpha$, X, and y are 44.0, 0.7, and 1.3, respectively. 
The $\alpha$ value corresponds to the case of low accretion rate in Figure 15 of \citet{2008ApJS..179..360G}, while the X value indicates the H-rich environment for the ignition of the burst. 
Additionally, the y value corresponds to the unstable burning regime ({\uppercase\expandafter{\romannumeral1}},{\uppercase\expandafter{\romannumeral2}}) in Figure 2 of \citet{2021ASSL..461..209G}, suggesting that the fuel of the burst is mixed H/He.

Based on the subgraph in Figure ~\ref{fig:figure5} and Figure~\ref{fig:figure8}, it is apparent that the majority of frequent long bursts are concentrated in the hard state, while most short bursts are prevalent in the soft state.
Interestingly, two PRE bursts were observed in the soft state.
Additionally, the burst profiles observed in the Intensity-Intensity diagram, as depicted in BAT and \textit{MAXI} data, align closely with the findings reported by \citet{2011MNRAS.413.1913Z} and \citet{2015MNRAS.454..541L}.

In the case of frequent long bursts, we found 30 bursts with similar profiles (Figure ~\ref{fig:figure5}). 
These bursts share common characteristics, including a roughly 4-second rise time, and a $\sim$ 75-second consistent decay time. 
These bursts occur predominantly in the hard state. 
These characteristics indicate that the bursts with similar profiles likely originate from comparable ignition conditions before the burst, and similar accretion environments during the burst.
Notably, this phenomenon has also been observed in GS 1826$-$24 during its hard state \citep{1989ESASP.296.....H,2004ApJ...601..466G}.
In that case, the average rise time of the bursts was 5.6$\pm$0.6 s and the average duration was over 150 s. 
In GS 1826$-$24, these regular bursts involve mixed hydrogen/helium (H/He) burning \citep{2000AIPC..522..359B} and undergo the rp process (i.e., Case 1 of \citet{1981ApJS...45..389W}). 
The duration of the burst in this source was longer than those observed in 4U 1636$-$53.  
It is possible that the higher hydrogen fraction in the fuel layer and the longer contribution time of the rp process are the reasons for this difference between 4U 1636$-$53 and GS 1826$-$24.
In the soft state, the bursts in 4U 1636$-$53 exhibited variations in rise times, peak rates, and durations.
Similarly, the bursts in GS 1826$-$24 displayed such variations.
These short bursts ignite fuel with significantly lower hydrogen fractions than the solar CNO metallicity \citep{2016ApJ...818..135C,2020A&A...634A..58S}.

\subsection{Two-stage decay phase}
\label{sec:Two-stage decay phase}

After fitting the decay phase of the bursts with two mathematical models, we found the decay phase can be divided in two parts (Figure~\ref{fig:figure5}), with the duration of the initial part being $\sim$ 6 s. 
This duration indicates that the cooling phase of the burst may exit the heating process and the domination component changed remarkably. 
Based on the explanation of the frequent long bursts in the hard state, the rp process may be responsible for the heating phase. 

The fact that the decay phase of the bursts cannot be described by a single exponential component has also been found in other sources. 
\citet{2017A&A...606A.130I} fitted the bolometric light curve of 1254 bursts from 60 sources with a model representing the rp process. 
The rp process component was described by the one-sided Gaussian function, derived from  a simplified model to describe the rp process in their work based on reaction chains described in \citet{1981ApJS...45..389W} and \citet{2001PhRvL..86.3471S}. 
They found that the decay phase has more than one stage, and the typical domination time of the rp process is $\sim$ 50 s in half of all bursts. 
We attempted to apply their model to our data, but our dataset could not effectively constrain the one-sided Gaussian component.
Their domination time of the rp process (50 s) is significantly longer than ours (6 s). 
Our initial duration is close to 5.658 s, which is the half-life of $^{21}{Mg}$ in the rp model.

The selection of the energy band may affect the measurement of the initial duration of the bursts. 
Ideally, the light curve should be measured over a broad energy band. 
However, obtaining the merged light curve over the full energy band of \textit{Insight}-HXMT is not feasible due to the different effective areas for LE and ME \citep{Zhang2020}.
Alternatively, the initial duration of the bursts can be measured using the bolometric light curve of the burst.
However, the errors of the parameters obtained after fitting the bolometric flux using the exponential model are relatively large. 
By overlapping the bolometric light curve and the 2$-$10 keV light curve on the same graph, we found that their trends are consistent. 
Therefore, the light curve observed only by the LE detector can be approximated to the bolometric flux variation of a burst.

\subsection{The persistent emission enhancement during the burst}
\label{the persistent emission enhancement during the burst}

We found that the $f_{\rm a}$ values of spectra in different categories of the bursts are mostly greater than 1 (Figure~\ref{fig:figure10}). 
This distribution indicates that the persistent emission has been enhanced during the burst. 
Some of the $f_{\rm a}$ values in PRE bursts are much larger than 1, indicating a substantial impact of the burst on the persistent emission, particularly during the burst's peak.
We found an inverse correlation between the persistent emission flux and the maximum ‘$f_{\rm a}$’ of all non-PRE bursts, as shown in the left plot of Figure~\ref{Figx:figure11}. 
Based on Section \ref{sec:Time Resolved Spectroscopy}, the persistent emission flux can be represented the accretion rate as a fraction of the Eddington rate. 
The X-axis of the left plot can be replaced with the accretion rate in the right plot. 
The anti-correlation suggests that during periods of lower accretion rates, there is a greater enhancement of the persistent emission near the burst peak. 

Since \citet{2013A&A...553A..83I} discovered the deviations between the spectral data and the single blackbody component during the burst in SAX J$1808.4-3658$, many explanations have been proposed.
One of the explanations is the reprocessing of the burst photon.
\citet{2013A&A...553A..83I} suggested that the increase of persistent emission during the burst may be the result of the reprocessing of the burst photon in the accretion disk.
Since no obvious reflection component was found in our spectrum, we cannot verify if reprocessing is relerant.
Another explanation was proposed by \citet{2017MNRAS.472...78K} based on the 4U 1608–52 soft PRE burst data.
They suggested that the increased area of the spreading layer contributes to additional persistent emissions during the burst.
In their interpretation, the spreading layer does not exist in the hard state, and there are only two soft PRE bursts in our data, which is not enough to verify their ideas.
The Poynting-Robertson drag and two comptonization components were also proposed to interpret $f_{\rm a}$ \citep{1989ApJ...346..844W,1992ApJ...385..642W,2013ApJ...772...94W,2005ApJ...634.1261T,2023JHEAp..40...76C}.
Next, we will discuss these explanations with our data.

\subsubsection{Poynting-Robertson drag}
\label{Poynting-Robertson drag}

The P-R effect is the radiation from the central source by losing and increasing the angular momentum of the surrounding accretion disk material, which in turn causes an inward or outward flow \citep{1989ApJ...346..844W}. 
The P-R drag is an inflow of the P-R effect.
Expanding upon this drag effect, \citet{1992ApJ...385..642W} predicted an anti-correlation between the pre-burst accretion rate and the enhancement of persistent emission near the burst peak.
This prediction is visually represented by the shaded region on the right side of Figure~\ref{Figx:figure11}. 
By comparison, the anti-correlation in the data is consistent with the prediction, but our data is generally below the predicted region. 
In the right panel of Figure~\ref{Figx:figure11}, we marked with the red diamond the values in Table 1 of \citet{1992ApJ...385..642W} that are close to the system parameters (the spin frequency) in 4U 1636$-$53.
If we used this point of 4U 1636$-$53 as a reference, the $f_{\rm a}$ values of our data are distributed around it. 
However, the accretion rates of our data are all about one order of magnitude lower than this point. 
This discrepancy may be attributed to our selection of an energy range for calculating the persistent emission flux (2$-$35 keV), while a portion of the emission from the accretion disk falls below 2 keV. 
This limitation measures that we underestimate the accretion rate.

In a prior study, \citet{2013ApJ...772...94W} found an anti-correlation between the persistent emission flux and the maximum $f_{\rm a}$ after an analysis of 332 PRE bursts observed with \textit{RXTE} in 40 sources.
Although the trend of PRE bursts in their observations aligned with the theoretical prediction of \citet{1992ApJ...385..642W}, their data were generally lower than the predicted maximum value of $f_{\rm a}$.
In their subsequent work, \citet{2015ApJ...801...60W}, they expanded their analysis to include 1759 bursts, encompassing both PRE and non-PRE bursts from 56 sources.
Their findings continued to support the existence of an anti-correlation.
The observed anti-correlation, both in their study and ours, is consistent with the theoretical prediction.
All this suggests that, during the burst, the radiation from the burst will induce the material on the accretion disk to accelerate toward the neutron star via the P-R drag, leading to an increase in the accretion rate and enhanced persistent emission.

\subsubsection{Two comptonization components}
\label{Two comptonization components}

 \begin{figure} 
	\includegraphics[width=\columnwidth]{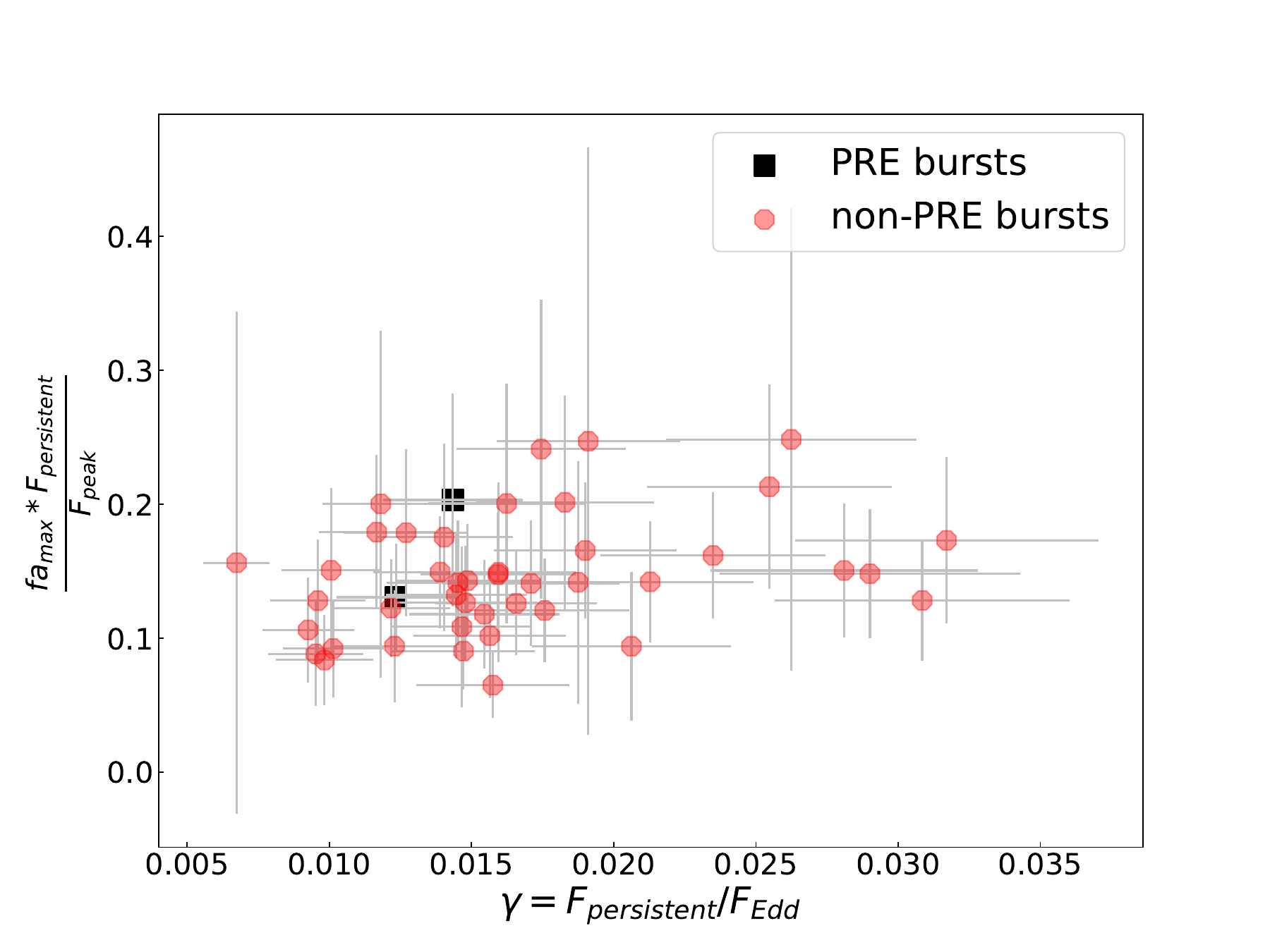}
    \caption{The relation between the comptonization-fraction and the accretion rate $\gamma$.
    }
    \label{fig:figure12}
\end{figure}

\citet{2005ApJ...634.1261T} used a two comptonization components model to describe the integrated persistent spectrum from GS 1826$-$238; one of the components comes from accretion disk corona (ADC) and the other one from, or near, the boundary layer (BL).
After the bursts start, only the corona near the BL changes, while the ADC remains constant.
In their fitting results, the photons from the burst would cool the inner corona from $\sim$ 6.8 keV to $\sim$ 3.0 keV near the burst peak, which contributes to the increase of the low energy emission during the burst.
In other words, the emission of the inner corona contributes to the persistent emission enhancement during the burst.
Evidence of the inner corona was also found in 4U 1636$-$53.
\citet{2017MNRAS.468.2256W} used the reflection model $RELXILLLP$ to fit the spectra observed by \textit{NuSTAR} and found that the height of the illuminator is relatively small, $\sim 2-3 R_g$.
This illuminator (inner corona zone) was interpreted as the BL.

%Chen
Based on the two comptonization components assumption near the burst peak, \citet{2023JHEAp..40...76C} defined the comptonization-fraction as ${f_{\rm a}}*F_{\rm persistent}/ F$, where $F$ is the bolometric flux of the burst, ${f_{\rm a}}*F_{\rm persistent}$ represents the energy from the photon scattering of the burst through the corona, and the ratio represents the fraction of photon energy enhancement during the burst.

Considering the physical basis for this fraction, and based on the theory of the P-R drag.
For an instantaneous luminosity $L$, we have $F=\xi L/4\pi R^2$ and taking into account the effect of geometric factor $\xi \sim max\{{R_*}/R, dH/dR\} $, where $R_*$ is the NS radius, $R$ is the radius of the disk, and $H$ is the disk thickness.
We could then write equation (2) of \citet{1989ApJ...346..844W} as:
\begin{equation}
\label{eq4}
   \dot{M}_{\rm PR} \sim \xi L / c^2
\end{equation}
for the mass transfer rate implied by the P-R drag torque and instantaneous luminosity $L$.
This mass transfer rate corresponds to the increase in the persistent emission, which is ${f_{\rm a}}*F_{\rm persistent}$.
If we divide both sides of the above equation by the peak luminosity $L$ of the burst, the left side corresponds to the comptonization-fraction and the right side is $ \xi / c^2$.
Consequently, the comptonization-fraction is proportion to $\xi$.
As the accretion rate increases, the edge of the inner disk moves towards the NS, the inner disk radius decreases, the geometric factor increases, and the comptonization-fraction increases.

During the burst, the emission from the neutron star is most pronounced near its peak and is therefore less affected by contamination from other processes.
Near the burst peak, the comptonization-fraction transforms to ${f_{\rm a}}_{\rm max}*F_{\rm persistent}/ F_{\rm peak}$.
% our work
To verify if the two comptonization components can explain the $f_{\rm a}$ in our data, we computed the relation between the comptonization-fraction and the accretion rate $\gamma$ in Figure ~\ref{fig:figure12}. 
This Figure shows that the comptonization-fraction is independent of $\gamma$, which is in line with the findings in 4U 1608$-$52 \citep{2023JHEAp..40...76C}.
Here, the fraction stabilizes at approximately 0.3, while $\gamma$ ranges from 0.03 to 0.18.
This indicates that the geometric factor remains unchanged as the accretion rate increases.
Notably, this pattern holds for PRE bursts as well, indicating the independence between the comptonization-fraction and the accretion rate.
This further proves that the fraction is not affected by burst intensity and accretion rate, and suggests that the maximum photospheric radius of the PRE burst may still not exceed the corona zone.
In conclusion, the impact of the P-R drag on the comptonization-fraction is negligible.
The enhancement of persistent emission is mainly contributed by the inner corona zone.

\section*{Acknowledgements}

J. M is supposed by the National Key R\&D Program of China (2023YFE0101200), Yunnan Revitalization Talent Support Program (YunLing Scholar Award), and the National Natural Science Foundation of China (NSFC) grant 11673062. 
GB acknowledges funding support from NSFC under grant Nos. U1838116.
Lyu is supported by Hunan Education Department Foundation (grant No. 21A0096). 
This work is supported by Nos. U2031205 and the National Key R\&D Program of China (2021YFA0718500).

%%%%%%%%%%%%%%%%%%%%%%%%%%%%%%%%%%%%%%%%%%%%%%%%%%
\section*{Data Availability}

The data for \textit{Insight}-HXMT underlying this article is available format at \textit{Insight}-HXMT website (http://archive.hxmt.cn/proposal; data in compressed format).



\bibliographystyle{mnras}
\bibliography{LMXB4U1636-53} % if your bibtex file is called example.bib

%\begin{thebibliography}{99}
%\end{thebibliography}

%%%%%%%%%%%%%%%%%%%%%%%%%%%%%%%%%%%%%%%%%%%%%%%%%%

%%%%%%%%%%%%%%%%% APPENDICES %%%%%%%%%%%%%%%%%%%%%

\appendix

\section{Some extra material}
%-------------------------------------------------------------------------
\begin{table*}
	\centering
	\caption{Overview of 45 X-ray bursts. 
Notes: ${*}$ indicates the ObsIDs out of the GTIs.
$\triangle$ indicates that the bursts have the HE data within the GTIs.
Burst \#26 has a non-exponential part in the decay phase, and its flux cannot be measured correctly. 
So the results of the decay phase in this burst will be neglected.
Burst \#2, \#3, \#4, \#12, \#15, and \#16 had been reported by \citet{2018ApJ...864L..30C}.}
	\label{tab:table2}
	\begin{tabular}{cccccccc}
		\hline
	   & & & & & Rise & e-folding & \\
        Burst & ObsID & Onset Date & Peak Flux & Fluence & Time & Time & $\tau$\\
          & & (MJD) & ($\times10^{-8}$erg s$^{-1}$cm$^{-2}$) & ($\times10^{-8}$erg cm$^{-2}$) & (s) & (s) & (s)\\
		\hline
        1$^{*}$ & P011465400201-20180211-01-01 & 58161.16566 & 5.36 $\pm$ 1.22 & 30.79 $\pm$ 2.82 & 1.25 & 3.16 & 5.75 \\  
        2 & P011465400301-20180213-01-01 & 58162.87104 & 5.85 $\pm$ 1.41 & 37.10 $\pm$ 2.85 & 1.75 & 3.98 & 6.35 \\  
        3$_\triangle$ & P011465400401-20180215-01-01 & 58164.73368 & 1.94 $\pm$ 0.63 & 21.55 $\pm$ 2.33 & 3.00 & 9.49 & 11.13 \\  
        4$_\triangle$ & P011465400501-20180217-01-01 & 58166.11714 & 1.96 $\pm$ 0.86 & 9.86 $\pm$ 1.56 & 2.25 & 3.74 & 5.04 \\ 
        5$^{*}$ & P011465400501-20180217-01-01 & 58166.20153 & 1.83 $\pm$ 0.82 & 11.46 $\pm$ 2.36 & 2.50 & 4.00 & 6.25 \\ 
        \hline
         & More information see Table-A1.txt & & & & &  \\
		\hline
	\end{tabular}
\end{table*}

%-------------------------------------------------------------------------

%-------------------------------------------------------------------------
\begin{table*}
	\centering
	\caption{Spectral parameters of 45 persistent 64s spectrum.}
	\label{tab:table3}
	\begin{tabular}{ccccccc}
		\hline
	   & & Norm & Flux & Flux & Flux & \\
        Burst & $\Gamma$ & (photons keV$^{-1}$s$^{-1}$cm$^{-2}$) & (2$-$35 keV) & (2-10 keV) & (10-35 keV) & $\chi_v^2$$/$dof\\
          & & (at 1 keV) & ($\times10^{-10}$erg s$^{-1}$cm$^{-2}$) & ($\times10^{-10}$erg cm$^{-2}$) & ($\times10^{-10}$erg s$^{-1}$cm$^{-2}$) &  \\
		\hline
        1 & 2.21 $\pm$ 0.12 & 0.39 $\pm$ 0.06 & 8.04 $\pm$ 0.31 & 7.37 $\pm$ 0.28 & 0.67 $\pm$ 0.03 & 0.89$/$66 \\ 
        2 & 2.15 $\pm$ 0.13 & 0.31 $\pm$ 0.05 & 6.92 $\pm$ 0.20 & 6.36 $\pm$ 0.19 & 0.57 $\pm$ 0.02 & 0.81$/$55 \\ 
        3 & 1.80 $\pm$ 0.10 & 0.20 $\pm$ 0.03 & 8.22 $\pm$ 0.23 & 7.50 $\pm$ 0.21 & 0.72 $\pm$ 0.02 & 0.80$/$63 \\ 
        4 & 2.03 $\pm$ 0.07 & 0.53 $\pm$ 0.05 & 14.71 $\pm$ 0.31 & 13.16 $\pm$ 0.28 & 1.55 $\pm$ 0.03 & 0.97$/$82 \\ 
        5 & 2.29 $\pm$ 0.09 & 0.58 $\pm$ 0.07 & 10.71 $\pm$ 0.33 & 9.88 $\pm$ 0.31 & 0.83 $\pm$ 0.03 & 0.87$/$69 \\ 
        \hline
         & More information see Table-A2.txt & & & & &  \\
		\hline
	\end{tabular}
\end{table*}
%-------------------------------------------------------------------------

%-------------------------------------------------------------------------
\begin{table*}
	\centering
	\caption{Spectral parameters of 45 bursts at the peak flux moment .}
	\label{tab:table4}
	\begin{tabular}{cccccccccc}
		\hline
          & & \multicolumn{2}{c}{Bolometric Flux} & \multicolumn{2}{c}{$kT_{\rm bb}$} & \multicolumn{2}{c}{${K}_{\rm bb}$} & \\
         Burst & $f_{\rm a}$ & \multicolumn{2}{c}{($\times10^{-8}$erg s$^{-1}$cm$^{-2}$)} & \multicolumn{2}{c}{(keV)} & \multicolumn{2}{c}{($R_{\rm km}^{2} / D_{10}^{2}$)} &  \multicolumn{2}{c}{$\chi_v^2$$/$dof}\\
          & & `$f_{\rm a}$' & `classical' & `$f_{\rm a}$' & `classical' & `$f_{\rm a}$' & `classical' & `$f_{\rm a}$' & `classical'\\
		\hline
        1 & 6.01 $\pm$ 1.86 & 4.73 $\pm$ 1.29 & 5.36 $\pm$ 1.22 & 2.63 $\pm$ 0.14 & 2.52 $\pm$ 0.12 & 91.79 $\pm$ 15.55 & 124.47 $\pm$ 16.05 & 0.75$/$46 & 0.95$/$47 \\ 
        2 & 4.35 $\pm$ 0.74 & 5.45 $\pm$ 1.39 & 5.85 $\pm$ 1.41 & 3.22 $\pm$ 0.17 & 3.07 $\pm$ 0.16 & 47.39 $\pm$ 6.57 & 61.44 $\pm$ 7.72 & 0.86$/$39 & 1.33$/$40 \\ 
        3 & 0.57 $\pm$ 0.59 & 1.96 $\pm$ 0.71 & 1.94 $\pm$ 0.63 & 1.88 $\pm$ 0.13 & 1.92 $\pm$ 0.13 & 146.36 $\pm$ 33.57 & 131.41 $\pm$ 25.13 & 0.99$/$20 & 0.93$/$21 \\ 
        4 & 2.44 $\pm$ 0.51 & 1.48 $\pm$ 0.95 & 1.96 $\pm$ 0.86 & 2.12 $\pm$ 0.27 & 2.22 $\pm$ 0.20 & 68.33 $\pm$ 26.41 & 75.39 $\pm$ 18.96 & 1.87$/$23 & 1.49$/$22 \\ 
        5 & 3.74 $\pm$ 1.38 & 1.63 $\pm$ 1.19 & 1.83 $\pm$ 0.82 & 2.12 $\pm$ 0.29 & 1.89 $\pm$ 0.17 & 74.61 $\pm$ 35.68 & 133.88 $\pm$ 36.14 & 0.97$/$18 & 1.12$/$19 \\ 
        \hline
         &  \multicolumn{2}{c}{More information see Table-A3.txt} & & & & & & & \\
		\hline
	\end{tabular}
\end{table*}

%%%%%%%%%%%%%%%%%%%%%%%%%%%%%%%%%%%%%%%%%%%%%%%%%%

% Don't change these lines
\bsp	% typesetting comment
\label{lastpage}
\end{document}